\documentclass[aps,amssymb,prr,twocolumn,superscriptaddress]{revtex4-1}
\usepackage{amsmath}
\usepackage[english]{babel}
\usepackage[utf8]{inputenc}
\usepackage{hyperref}
\usepackage{float}
\usepackage{graphicx}
\usepackage{braket}
\usepackage{xcolor}

\allowdisplaybreaks

\begin{document}
	\title{Tunneling conductance of long-range Coulomb interacting Luttinger liquid}
	
	\author{DinhDuy Vu}
	\affiliation{Condensed Matter Theory Center and Joint Quantum Institute, Department of Physics, University of Maryland, College Park, Maryland 20742, USA}
	
	\author{An\'{\i}bal Iucci}
	\affiliation{Condensed Matter Theory Center and Joint Quantum Institute, Department of Physics, University of Maryland, College Park, Maryland 20742, USA}
	\affiliation{Instituto de F\'{\i}sica La Plata - CONICET, Diag 113 y 64 (1900) La Plata, Argentina}
	\affiliation{Departamento de F\'{\i}sica, Universidad Nacional de La Plata, cc 67, 1900 La Plata, Argentina.}
	
	\author{S. Das Sarma}
	\affiliation{Condensed Matter Theory Center and Joint Quantum Institute, Department of Physics, University of Maryland, College Park, Maryland 20742, USA}
	
	\begin{abstract}
		The theoretical model of the short-range interacting Luttinger liquid predicts a power-law scaling of the density of states and the momentum distribution function around the Fermi surface, which can be readily tested through tunneling experiments. However, some physical systems have long-range interaction, most notably the Coulomb interaction, leading to significantly different behaviors from the short-range interacting system. In this paper, we revisit the tunneling theory for the one-dimensional electrons interacting via the long-range Coulomb force. We show that even though in a small dynamic range of temperature and bias voltage, the tunneling conductance may appear to have a power-law decay similar to short-range interacting systems, the effective exponent is scale-dependent and slowly increases with decreasing energy. This factor may lead to the sample-to-sample variation in the measured tunneling exponents. We also discuss the crossover to a free Fermi gas at high energy and the effect of the finite size. Our work demonstrates that experimental tunneling measurements in one-dimensional electron systems should be interpreted with great caution when the system is a Coulomb Luttinger liquid.
	\end{abstract}
	\maketitle
	\section{Introduction}
	Luttinger liquids emerge from interacting one-dimensional many-electron systems where the Fermi surface is two discrete points rather than a connected surface as in higher-dimensional cases. Two and three-dimensional interacting systems are conventionally described by the Fermi liquid model where the excitations are individual quasi-particles with renormalized properties (e.g. effective mass) from the bare particle. Fermi liquids manifest qualitative resemblance to the free Fermi gas, for example, the discontinuity of the momentum distribution through the Fermi momentum and singularities in the spectral function representing quasi-particles. However, in one-dimensional systems, quasi-particle excitations are replaced by collective excitations even for very weak interaction, leading to the complete disappearance of the one-to-one correspondence with the non-interacting Fermi gas. Most remarkably, the momentum distribution function is continuous through the Fermi point and the density of states displays a pseudo gap at the Fermi energy. These features indicate the breakdown of the quasi-particle picture. More non-Fermi liquid phenomenon include charge-spin separation, and power-law scaling of charge and spin correlations  \cite{Haldane1981a,Voit1995,Voit1993,Giamarchi2004}. Assuming a zero-range (or short-range) interaction (e.g. a Dirac delta function), the density of states decays as $(E-E_F)^\alpha$ where $\alpha$ is a finite and nonuniversal constant since it depends on the actual interaction strength. In addition, this constant $\alpha$ also shows up in the tunneling conductance, giving it a distinct power-law decaying behavior. Specifically, if the tunneling is between a Luttinger liquid characterized by the exponent $\alpha$ and an ordinary Fermi-liquid metal, the tunneling conductance at temperature $T$ and bias voltage $V_0$ is $G=dI/dV_0 \propto V_0^{\alpha}$ for $eV_0\gg k_BT$ and $G\propto T^{\alpha}$ for $eV_0\ll k_BT$. For tunneling between two Luttinger liquids, the exponent is simply doubled \cite{Kane1992,Kane1992a,Chang2003}. This power-law tunneling behavior is considered a signature of the Luttinger liquid since in Fermi liquids, $G$ is simply a constant for small values of $T$ and $V_0$ (as long as $k_BT,eV_0\ll E_F$, where $E_F$ is the Fermi energy). Indeed, tunneling experiments have confirmed this behavior in many physical systems. Earliest attempts include chiral Luttinger liquids found in the edge mode of fractional quantum Hall fluids \cite{Chang2003,Wen1991,Chang1996,Moon1993} and the power law in the optical response from quasi-one-dimensional conductors \cite{Dardel1991,Dudy2012}, suggesting the Luttinger liquid nature of these systems. Recently, studies have focused on carbon nanotubes where extreme isolating conditions can be obtained to create a strongly correlated non-chiral 1D electron system. Tunneling experiments on carbon nanotubes also show evidences for power laws characterizing Luttinger liquids \cite{Yao1999,Postma2000,Zhao2018}.

	It is instructive to review the theoretical description of the Luttinger liquid based on the bosonization method, which in principle can give exact solutions for any types of interactions provided that the back-scattering is ignored and the dispersion can be linearized.  If we assume that the system has only contact charge-charge interaction with a strength $U$, i.e. $H_{int} = U/2\int \rho(x)^2 dx$; the plasmon velocity is renormalized by $g= v_F/v_\rho = (1+NU/(\pi v_F))^{-1/2}$ where $N$ is system degeneracy (e.g. $N=2$ for spinful systems or $N=4$ for carbon nanotubes including spin and valley without spin-orbit coupling). The exponent of the density of states is then $\alpha_{\mathrm{bulk}}=(g+g^{-1}-2)/2N$ at the bulk of the system and $\alpha_{\mathrm{bound}}=(g^{-1}-1)/N$ at the open boundary \cite{boundary1,boundary2}. For systems with repulsive interaction, $U>0$, so $g<1$ and $\alpha_{\mathrm{bound}} > \alpha_{\mathrm{bulk}} >0$. 
	
	The simplicity of the short-range interacting Luttinger liquid model might be attractive, but the model of zero-range interaction is only an approximation. Most real systems have long-range interactions, most notably the electronic Coulomb interaction. If the Fourier transform of the interaction approaches a finite constant at the zero transfer momentum limit, the low-energy physics can be described by a short-range interaction model without great loss of accuracy. However, the Coulomb interaction is special because of its logarithmic divergence at small momentum. One way to obtain a short-range interaction is to consider an appropriately screened Coulomb interaction \cite{kane,screenCoulomb}, imagining the carbon nanotube is placed inside a larger metallic tube with a radius $R_s$ much larger than the radius $R$ of the nanotube . Then, the interaction potential is totally screened out and can be considered as the classical energy of the E-field trapped between the nanotube and the metallic cylinder $H_{int}= e^2\ln (R_s/R)\int \rho(x)^2 dx$. This is exactly the form of the zero-range interaction, so even if the nanotube has long-range interaction, the Luttinger liquid considerations from the short-range case still apply with an appropriate $g$ provided, of course, the nanotube is indeed enclosed in a metallic cage. However, the ambiguity here is the value $R_s$ since in reality, no specific metal tube encloses the nanotube.  Note that the logarithmic divergence of the effective interaction in $H_{int}$ as $R_s$ goes to infinity is the well-known logarithmic divergence of the 1D Coulomb interaction arising from its long-range nature, which is simply cut off by taking $R_s$ to be finite.
	
	A more rigorous approach is not to make an ad hoc short-range interaction approximation, and instead use the long-range 1D Coulomb interaction itself in the Luttinger liquid theory, i.e. performing bosonization with the logarithmically divergent interaction originating from the 1D $1/x$ Coulomb interaction. Within this description, the power law behavior is no longer valid. Indeed, for the density-density correlation, the $4k_F$ oscillation decays as $x^{-\sqrt{\ln x}}$, slower than any power law \cite{Schulz,Coulomb_Luttinger2}. The 1D Coulomb Luttinger liquid is thus an effective Wigner crystal at finite length with the $4k_F$ density oscillation decaying very slowly over distance (since the correlation dies out eventually there cannot be any real long-range order). Moreover, due to the log-divergence, the ``effective exponent" of a Coulomb Luttinger liquid is scale-dependent, i.e. $\alpha$ is also a function of energy \cite{Coulomb_Luttinger1}. This scale-dependence of the effective Luttinger exponent in the Coulomb Luttinger liquid, in contrast to the constant exponent for the short-range interaction model, complicates the description of tunneling measurement as the effective exponents also depend on the temperature or voltage scale being studied. It is unclear apriori how one can investigate the Coulomb Luttinger liquid simply by mapping it into a corresponding short-range interaction model although this is often done in the interpretation of the experimental 1D results.
	
	In this paper, we investigate the effect of the long-range Coulomb interaction on the tunneling conductance using the long-range interaction description of a scale-dependent exponent. One specific consequence of the scale-dependent exponent of the Coulomb Luttinger liquid may be the manifestation of the sample-to-sample variations in the measured tunneling exponent often seen within even a given type of 1D physical systems.  Since the Coulomb Luttinger liquid by definition does not have a constant exponent (i.e. the exponent varies slowly over the energy scale of measurements), it is important to analyze the tunneling experiment in depth using the long-range interaction model to figure out how this scale dependence might manifest in the tunneling spectroscopy.  This is the main goal of the current work.  A secondary goal is to investigate the role of the finite length of the 1D system (e.g. carbon nanotube or semiconductor quantum wire) in the tunneling experiments to check whether an implicit or explicit length dependence affects the tunneling exponent, particularly in the context of the scale-dependent exponent  in the Coulomb Luttinger liquid. 
	
	The rest of the paper is organized as follows.  In Sec. II we provide the theoretical description for a Coulomb Luttinger liquid with the open boundary condition and compare its properties with the short-range interacting Luttinger liquid counterpart. We also discuss the crossover to the free Fermi gas behavior at high energy, which may not be obvious in the short-range model but appears naturally in the long-range Coulomb Luttinger liquid. In Sec. III, we study the effect of the finite system, finding it to be irrelevant as long as the 1D system is not too short. We conclude in Sec. IV summarizing our main findings and discussing possible experimental implications of our results.
	
	\section{Theoretical model}
	
	\subsection{Bosonization in open-boundary systems}
	We first present the bosonization study of a 1D $N-$fold degenerate system of size $L$ and open boundaries at $x=0$ and $x=L$ which are appropriate for tunneling measurements. Due to the open boundary condition, the left and right-moving electrons are no longer independent operators. Accordingly, the fermion field can be decomposed as
	\begin{equation}
	\begin{split}
	  \psi_s(x) &= \psi_{L,s}(x)+\psi_{R,s}(x) \\
	          &= \frac{i}{\sqrt{2L}}\sum_{k}e^{-ikx}c_{k,s} - \frac{i}{\sqrt{2L}}\sum_{k}e^{ikx}c_{k,s},
	\end{split}
	\end{equation}
	where $s\in\{s_1,s_2,\dots,s_N\}$ is the electron species (e.g spin, valley, etc.) index. The construction implies that $\psi_{R,s}(x)=-\psi_{L,s}(-x)$. The linearized Hamiltonian with only charge-charge interaction in the unit system of $\hbar=1$ is
	\begin{equation}
	 H= v_F\int\sum_s \psi_s^\dagger\frac{i\partial\psi_s}{\partial x} dx + \int \frac{  U(x-y) \rho(x)\rho(y)}{2}dxdy,
	\end{equation}
	where $\rho(x)=\sum_s \rho_s(x)$ is the charge density. It is noted that for an open boundary system, the discretized momentum is $k=n\pi/L$ instead of $2\pi/L$ as in the periodic one. We define bosonic creation and annihilation operators with $q=m\pi/L>0$ as
	\begin{equation}
	\begin{split}
	 & a^\dagger_{q,s}=\sqrt{\frac{\pi}{qL}}\sum_{k} c_{k+q,s}^\dagger c_{k,s}\\ 
	 & a_{q,s}=\sqrt{\frac{\pi}{qL}}\sum_{k} c_{k-q,s}^\dagger c_{k,s},
	\end{split}
	\end{equation}
	and $[a_{q,s},a^\dagger_{q',s'}]=\delta_{q,q'}\delta_{s,s'}$. The right-moving chiral fermion field is
	\begin{equation}
	\begin{split}
	 	&\psi_{R,s}(x)=\lim\limits_{\epsilon\to 0^+}\frac{-i}{\sqrt{2}}\frac{F_s}{\sqrt{2\pi\epsilon}}e^{\frac{i\pi x N_s}{L}} e^{\phi_s(x)}; \text{ where }\\
	 	&\phi_s(x)= \sum_{q} \sqrt{\frac{\pi}{qL}}e^{-\epsilon q/2}\left(e^{iqx}a_{q,s} -e^{-iqx}a^\dagger_{q,s} \right),	
	\end{split}
	\end{equation}
	$N_s$ is the number operator of the $s-$electrons. $F_s$ is an operator such that $F_s$ commutes with all bosonic operators, $F_s\ket{N_s}=\ket{N_s-1}$ and $F_s^\dagger\ket{N_s}=\ket{N_s+1}$. One can perform a unitary transformation to separate the charge channel from other channels
	\begin{equation}
		a'_{q,n} = \frac{1}{\sqrt{N}}\sum_{m=1}^N e^{\frac{2i\pi mn}{N}} a_{q,s_m},
	\end{equation}
	The Hamiltonian is then the sum of separated channels $H=\sum_{n=0}^{N-1} H_n$. The interacting charge/plasmon channel ($n=0$) is described by the Hamiltonian
	\begin{equation}
	\begin{split}
	H_0 = &\frac{\pi v_0 |N_0|^2}{2L} +\sum_{q>0}q\left[v_F+\frac{NU(q)}{2\pi}\right]a'^\dagger_{q,0}a'_{q,0},\\ &\quad
	 +q\frac{NU(q)}{4\pi}\left(a'^\dagger_{q,0}a'^\dagger_{q,0} + a'_{q,0}a'_{q,0}\right),
	\end{split}
	\end{equation}
	where $v_0=v_F+NU(0)/2\pi$, and $U(q)=\int_{-L}^{L} U(x)e^{iqx} dx$. For the other non-interacting channels
	($n=1,\dots,N-1$),
	\begin{equation}
		H_n = \frac{\pi v_F |N_n|^2}{2L} + \sum_{q}v_F a'^\dagger_{q,n}a'_{q,n}.
	\end{equation}
	After the Bogoliubov transformation, the Luttinger interaction parameter is defined as
	\begin{equation}\label{eq6}
	g(q) = e^{2\theta} = \left[\frac{v_F}{v_F+NU(q)/\pi}\right]^{1/2},
	\end{equation}
	and the collective plasmon mode velocity is
	\begin{equation}
		v_\rho(q)=\frac{v_F}{g(q)} = v_F\sqrt{1+\frac{NU(q)}{\pi v_F}}.
	\end{equation} 
	For a short-range interaction, $U(q)$ is a constant and hence the constancy of $g$ as a Luttinger exponent for a given model, i.e., a given $U$. For a long-range interaction, $U(q)$ is obviously scale-dependent as it depends explicitly on the momentum $q$, resulting in the remarkable divergence $g(q)\sim \ln(1/q)$ in the case of 1D Coulomb interaction. The fermion correlation function is 
    \begin{equation}
    \begin{split}
       &\braket{\psi_s(x,t)\psi_s^\dagger(y,0)} = 2\braket{\psi_{R,s}(x,t)\psi_{R,s}^\dagger(y,0)}\\
      &-\braket{\psi_{R,s}(x,t)\psi_{R,s}^\dagger(-y,0)}-\braket{\psi_{R,s}(-x,t)\psi_{R,s}^\dagger(y,0)}.
    \end{split}
    \end{equation} Assuming $\braket{N_s}$=$N_e/N$ with $N_e$ being the total number of electrons, the chiral correlation is
	\begin{equation}
	\braket{\psi_{R,s}(x,t)\psi_{R,s}^\dagger(y,0)}=e^{ik_F(x-y)-iE_Ft}C(x,y;t);
	\end{equation}
	where
	\begin{equation}\label{eq7}
      \begin{split}
      &C(x,y;t) =\lim \limits_{\epsilon\to 0^+}\frac{e^{\frac{i\pi }{2L}(x-y-v_ct)}}{4\pi\epsilon}\\
      & \times \exp\left[\frac{F(x,y;t)+iK(x,y;t)}{N}\right]\left[\frac{\pi t }{\beta\sinh(\pi t/\beta) }\right]^{\frac{N-1}{N}}.
      \end{split}
    \end{equation}
	Here, we define the Fermi momentum $k_F=(N_e/N+1/2)\pi/L$, the Fermi energy $E_F=v_0N_e\pi/(NL)$ and the charge gap velocity $v_c=[v_0+(N-1)v_F]/N$ (i.e. the average over all channels). For brevity, we drop the $q$ argument in $v_\rho(q)$ and $g(q)$, and use $\cosh\theta=(g^{1/2}+g^{-1/2})/2$ and $\sinh\theta=(g^{1/2}-g^{-1/2})/2$. We have
	\begin{widetext}
		\begin{equation}\label{eq8}
		\begin{split}
		K(x,y;t) & =\sum_{q>0}e^{-\epsilon q} \left(\frac{\pi}{qL}\right)\left[ \cosh^2\theta\sin q(x-y-v_\rho t)-\sinh^2\theta\sin q(x-y+v_\rho t) \right. \\
		& \left. \quad -\sinh 2\theta \sin q(x+y-v_\rho t)/2+\sinh 2\theta \sin q(x+y+v_\rho t)/2 \right],\\
		F(x,y;t) &=-\sum_{q>0} e^{-\epsilon q} \left(\frac{\pi}{qL}\right)[1+2f_B(v_\rho q)]\times [\cosh^2\theta(1-\cos q(x-y-v_\rho t))+\sinh^2\theta(1-\cos q(x-y+v_\rho t))\\
		&-\sinh 2\theta(\cos 2qx+\cos 2qy-\cos q(x+y-v_\rho t)-\cos q(x+y+v_\rho t))/2],
		\end{split}
		\end{equation}		
	\end{widetext}
	where $f_B(z) = \left(e^{\beta z}-1\right)^{-1}$ is the Bose-Einstein distribution coming from the bosonic plasmon with $z$ being the excitation energy and $\beta=1/k_BT$ being the inverse electron temperature. Equations~\eqref{eq7} and \eqref{eq8} are the exact expressions and do not assume any specific form of the interaction.
	
	We can perform a quick check on the power law for $T=0$, $L\to \infty$ (in that case $\sum_q \pi/L \to \int dk$) and zero-range interaction $g(q)=g$. The spinless ($N=1$) dynamic chiral correlation function is
	\begin{equation}
	C(x,x;t)\propto \left(\frac{1}{t}\right)^{\frac{g+g^{-1}}{2}}\left( \frac{4x^2}{|4x^2-v_\rho^2t^2|} \right)^{\frac{g^{-1}-g}{4}}.
	\end{equation}
	This term is the primary contribution to the correlation function because the other term $C(x,-x;t)$ has fast $2k_F$ oscillation of $e^{2ik_Fx}$ and is further suppressed by $x^{-g}$. For $v_\rho t \gg x$ (near the boundary), $C(x,x;t) \propto t^{-g^{-1}}$ corresponding to the density of states $\rho(\omega) \propto \omega^{g^{-1}-1}$ ($\omega=E-E_F$) for $\omega \ll v_\rho /x$. For $v_\rho t \ll x$ (far from the boundary), $C(x,x;t) \propto t^{-(g^{-1}+g-2)/2}$. As a result, $\rho(\omega)\propto \omega^{(g^{-1}+g-2)/2}$ for $\omega \gg v_\rho/x$. These are of course well-known results provided here for the sake of completeness and to set a context for our work.	
	
	\subsection{Coulomb Luttinger effective exponent}
	In this section, we assume the semi-infinite 1D limit $L\to \infty$ (the effect of finite $L$ is discussed later). We study two models of interactions in the four-fold degenerate carbon nanotube system: (i) short-range interaction with constant $g$ up to a cut-off $\Lambda$, i.e.
	\begin{equation}
	g(q) = \begin{cases}
	& g  \text{ for } q\le\Lambda\\
	& 0  \text{ for } q>\Lambda
	\end{cases}.
	\end{equation}
	As long as $\omega<E_0 = \Lambda v_F$, the density of states $\rho(\omega) \propto \omega^{(g^{-1}-1)/4}$ at the boundary and $\rho(\omega)\propto \omega^{(g^{-1}+g-2)/8}$ in the bulk. It is noted that the cut-off $\Lambda$ is purely artificial, in fact, many theoretical works only consider the large distance asymptote and set $\Lambda\approx 1/\epsilon \to \infty$. (ii) The 1D Coulomb interaction $U(x)=e^2/(\kappa\sqrt{x^2+d^2})$ where $\kappa$ is the dielectric constant and $d$ is proportional to the transverse size of the nanotube, which regularizes the 1D Coulomb coupling. The corresponding $g$ is 
	\begin{equation}\label{longrange}
		g(q) = \left[1+\frac{8e^2}{\kappa\pi v_F} K_0(qd)\right]^{-1/2},
	\end{equation}
	with $v_F=8\times 10^5$~m/s and $K_0$ is the Bessel function. Note that in Ref. \cite{kane}, the Coulomb Luttinger liquid is approximated by an effective short-range interaction model with a constant $g$ as
	\begin{equation}
	g=\left[ 1+ \frac{8e^2}{\pi v_F} \ln\left(\frac{R_s}{R}\right) \right]^{-1/2},
	\end{equation}
	where $R$ is the radius of the tube and $R_s$ is some screening length; for $R_s \sim 100 R$, $g \sim 0.2$. We emphasize that $R_s$ is unknown and arbitrary in the experimental systems, and often used simply as a fitting parameter uncritically.
	
	We begin by studying the tunneling density of states in the bulk and the boundary of a Coulomb Luttinger liquid. The following argument is based on Wang et al. \cite{Coulomb_Luttinger1}, and extended to include the open boundary condition of relevance to tunneling spectroscopy. The bulk density of states in a carbon nanotube is given by
	\begin{equation}
	\begin{split}
 	&\rho_{\mathrm{bulk}}(\omega) = \frac{1}{\pi}\int_{0}^{\infty} \frac{1-\cos\omega t}{v_Ft}\text{Im}\left(e^{-\Phi_{\mathrm{bulk}}}\right)dt ;\\
 &\Phi_{\mathrm{bulk}}(t)=\int_{0}^{\infty} \frac{dq}{4q}\left[\left(1-e^{iv_\rho qt}\right)\frac{g^{-1}+g}{2} - \left(1-e^{iv_Fqt}\right) \right].
	\end{split}
	\end{equation}
	In the phase factor $\Phi_{\mathrm{bulk}}(t)$, when $q<A/t$ with $A$ is some number, the factor $1-e^{iqvt} \approx 0$ and when $q>A/t$, this factor is fast oscillating and negligible, thus the leading order term in the phase factor is
	\begin{equation}
	\begin{split}
	&\Phi_{\mathrm{bulk}}(t)  \approx \frac{1}{4} \int_{A/t}^{\infty} \frac{g(q)+g(q)^{-1}-2}{2q}dq \\
	& \approx \frac{\sqrt{U_0}}{12}\ln^{3/2}\left(\frac{q_st}{A}\right) + \frac{1}{4\sqrt{U_0}}\ln^{1/2}\left(\frac{q_st}{A}\right) - \frac{1}{4}\ln\left(\frac{q_st}{A}\right),
	\end{split}
	\end{equation}
	where we have used the low-$q$ asymptotic form $g(q) = \sqrt{U_0}\ln^{1/2}(q_s/q)$ with $U_0=8e^2/(\kappa \pi v_F)$ and $q_s\approx 1.12e^{1/U_0}/d$ in the integration instead of the full form Eq.~\eqref{longrange}. The power law of the density of states is then
	\begin{equation}
	\begin{split}
	&\rho_{\mathrm{bulk}}(\omega) \approx \left(\frac{\omega}{\omega_s}\right)^{\gamma_{\mathrm{bulk}}(\omega)};\text{ where}\\
	&\gamma_{\mathrm{bulk}}(\omega) = \frac{\sqrt{U_0}}{12}\ln^{1/2}\left(\frac{\omega_s}{\omega}\right) + \frac{1}{4\sqrt{U_0}}\ln^{-1/2}\left(\frac{\omega_s}{\omega}\right) - \frac{1}{4},
	\end{split}
	\end{equation}
	and the scale $\omega_s=20 v_Fq_s\sqrt{U_0}$. Extending the argument to the limit $x\to 0$, the boundary phase factor is
	\begin{equation}
	\begin{split}
	\Phi_{\mathrm{bound}}(t) & \approx \frac{1}{4} \int_{A/t}^{\infty} \frac{g(q)^{-1}-1}{q}dq\\
	&\approx \frac{\sqrt{U_0}}{6}\ln^{3/2}\left(\frac{q_st}{A}\right) - \frac{1}{4}\ln\left(\frac{q_st}{A}\right).
	\end{split}
	\end{equation}
	Accordingly, the local density of states at the open boundary is 
	\begin{equation}
	\begin{split}
	&\rho_{\text{bound}}(\omega)\approx \left(\frac{\omega}{\omega_s}\right)^{\gamma_{\text{bound}}(\omega)}, \text{ where}\\
	&\gamma_{\mathrm{bound}}(\omega) = \frac{\sqrt{U_0}}{6}\ln^{1/2}\left(\frac{\omega_s}{\omega}\right) - \frac{1}{4}.
	\end{split}
	\end{equation}
	As $\gamma$ depends on the energy $\omega$, strictly speaking this is not a power law. However, we can define an effective exponent $\alpha = d\ln(\rho(\omega))/d\ln\omega$ for the bulk and boundary tunneling density of states as
	\begin{equation}\label{eq24}
	\begin{split}
	& \alpha_{\mathrm{bulk}} = \frac{\sqrt{U_0}}{8}\ln^{1/2}\left(\frac{\omega_s}{\omega}\right) + \frac{1}{8\sqrt{U_0}}\ln^{-1/2}\left(\frac{\omega_s}{\omega}\right) - \frac{1}{4},\\
	& \alpha_{\mathrm{bound}} = \frac{\sqrt{U_0}}{4}\ln^{1/2}\left(\frac{\omega_s}{\omega}\right)- \frac{1}{4}.
	\end{split}
	\end{equation}
	We expect these exponents to appear in the tunneling conductance of the Coulomb Luttinger liquid. We note that the energy dependence of $\alpha$ as reflected in the explicit appearance of $\omega$ in the right hand side of Eq.~\eqref{eq24} leads to an ill-defined scale-dependent exponent in the Coulomb Luttinger liquid in contrast to the constant (but nonuniversal) exponent in the short-range interaction model.
	
	\subsection{Tunneling conductance}
	Suppose electrons tunnel between systems 1 and 2 at voltage bias $V_0$, the tunneling Hamiltonian is given by \cite{tunneling,Kane1992,Chang2003}
	\begin{equation}
		H_{tunnel} \sum_{s,s'} \Gamma_{s,s'} (\psi_{1,s}^\dagger(x)\psi_{2,s'}(x)+h.c.),
	\end{equation}
	where $\psi_{1,s}$ and $\psi_{2,s'}$ correspond to the $s$ and $s'-$electron wavefunction in the system 1 and 2 respectively. Our tunneling Hamiltonian corresponds to a perturbative point-contact, which is a common experimental setup. One can refer to other works for more rigorous tunneling conditions, for instance, non-perturbative contact \cite{HOtunnel}, back-scattering defects \cite{back_scat} or a $Y-$junction of three nanowires \cite{Yjunction1,Yjunction2,Yjunction3,Yjunction5,Yjunction6,Yjunction7}. Within the scope of this paper, the electron species degeneracy reduces the problem to the tunneling between two single-mode Luttinger liquids because the electron correlation function is identical for all electron species (see Eq.~\ref{eq7} and \eqref{eq8}). For our purpose, therefore, we do not need to worry about the symmetry or degeneracy aspects of the Luttinger liquid. The tunneling current is then given by
	\begin{equation}\label{tunnel}
	\begin{split}
		I \propto \int e^{ieV_0t} \text{Im}\left[ C_1(x,x;t) C_2(x,x;t)\right] dt.
	\end{split}
	\end{equation}
	Equation \eqref{tunnel} is basically the Fermi golden rule expressed in the Fourier transform. We specify the differential tunneling conductance $G=dI/dV_0$ in two distinct experimental setups: the two systems 1 and 2 are two identical Luttinger liquids, denoted as L-L,
	\begin{equation}\label{eq14}
	\begin{split}
		G_{L-L} &\propto  1-2\int_{0}^{\infty}\frac{\cos(eV_0t)}{t} \left[\frac{\pi t }{\beta\sinh\left(\pi t/\beta \right)}\right]^{2-\frac{2}{N}}\\
		&\times \exp\left[\frac{2F(x,x;t)}{N}\right]\sin\left[\frac{2K(x,x;t)}{N}\right]dt;
	\end{split}
	\end{equation}
	 and, system 1 is the Luttinger liquid sample and system 2 is a conventional 3D Fermi liquid metal contact, denoted as L-M, with $C_2(x,x;t)\sim 1/t$ (assuming the temperature is much less than the metal Fermi temperature)
	\begin{equation}\label{eq15}
	\begin{split}
	G_{L-M} &\propto  1-2\int_{0}^{\infty}\frac{\cos (eV_0t)}{t} \left[\frac{\pi t }{\beta\sinh\left(\pi t/\beta \right)}\right]^{1-\frac{1}{N}}\\
	&\times \exp\left[\frac{F(x,x;t)}{N}\right]\sin\left[\frac{K(x,x;t)}{N}\right]dt.
	\end{split}
	\end{equation}	
	From Eqs.~\eqref{eq14} and \eqref{eq15}, we can expect that the exponent of the L-L tunneling is two times as large as that of the tunneling through L-M contact. We recall that $F(x,x;t)$ and $K(x,x;t)$ are given in Eq.~\eqref{eq8}.

\begin{figure}
    	\begin{minipage}{0.02\textwidth}
    		\rotatebox{90}{\hspace{0.4in}$d\ln G/d\ln(V_0)$}
    	\end{minipage}
    	\begin{minipage}{0.45\textwidth}
    		\centering
    		\includegraphics[width=0.95\textwidth]{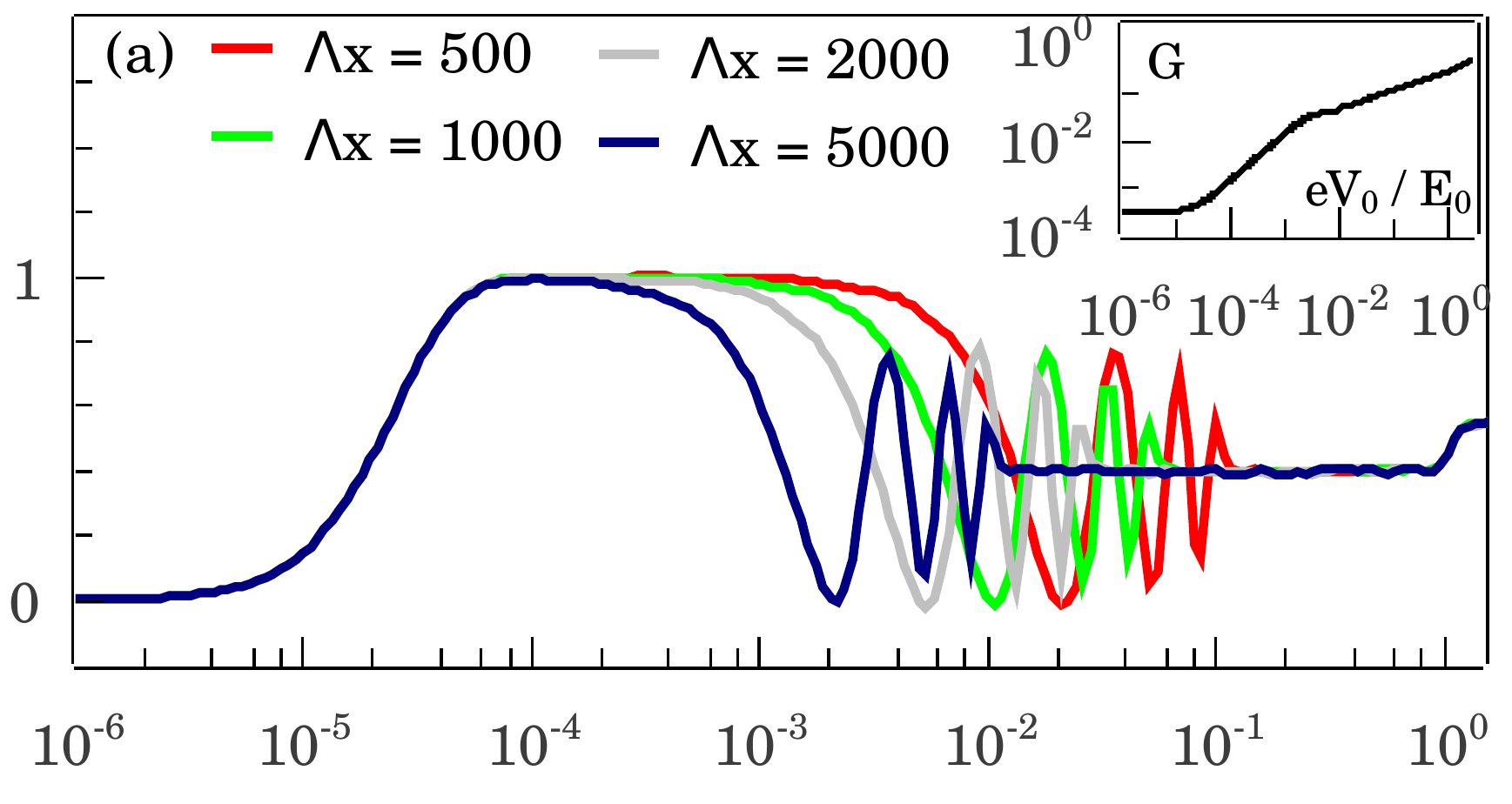}
    		$eV_0/E_0$  	
    		\vspace{0.1in} 		
    	\end{minipage}
    	\begin{minipage}{0.02\textwidth}
    		\rotatebox{90}{\hspace{0.2in}$d\ln G/d\ln(T)$}
    	\end{minipage}
    	\begin{minipage}{0.45\textwidth}
    		\centering
    		\includegraphics[width=0.95\textwidth]{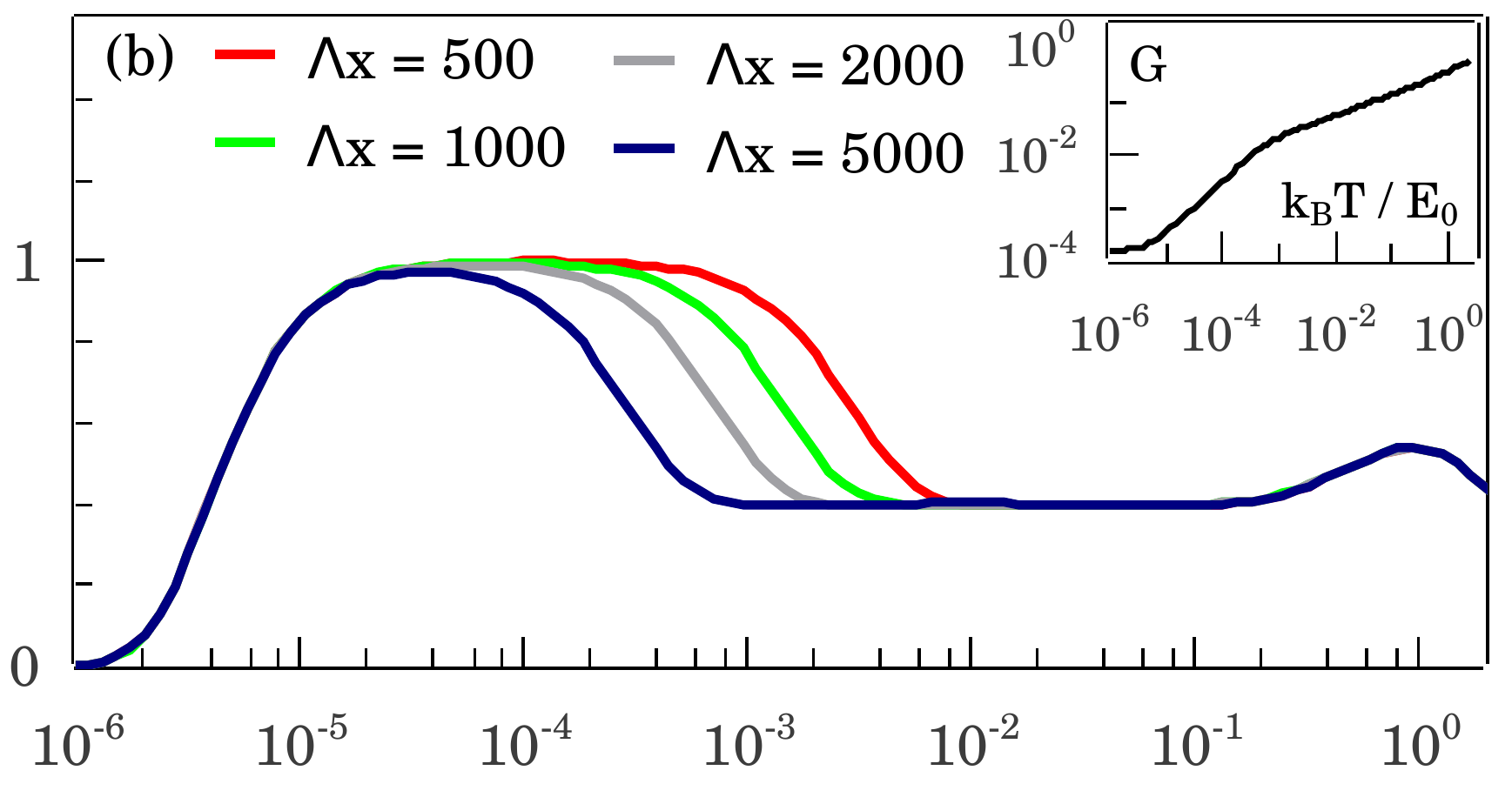}
    		$k_BT/E_0$  	 		
    	\end{minipage}			
    	\caption{The effective exponent of the L-M tunneling conductance for a short-range interacting Luttinger liquid with $g=0.2$ (a) with respect to $eV_0/E_0$ at fixed temperature $k_BT=10^{-5}E_0$ and (b) with respect to $k_BT/E_0$ at fixed voltage bias $eV_0=10^{-5}E_0$. The energy scale is $E_0=v_F\Lambda$. The insets show the tunneling conductance in arbitrary unit for $\Lambda x=2000$.}\label{fig1}
    \end{figure}

	In Fig.~\ref{fig1}, we show the calculated L-M tunneling conductance of a short-range interacting Luttinger liquid with $g=0.2$ in the inset and the effective exponent in the main plot as a function of voltage bias and temperature. There are three energy scales in the plot: the saturation regime ($eV_0\ll k_BT$ or $k_BT\ll eV_0$) where $G$ is independent of $V_0$ (or $T$), thus the effective exponent approaches zero, the boundary regime ($eV_0, k_BT\ll  v_\rho/x$) where the power law is given by the boundary relation $\alpha=(g^{-1}-1)/4 =1.0$ and the bulk regime ($eV_0, k_BT \gg  v_\rho/x$) where the power law is given by the bulk relation $\alpha=(g+g^{-1}-2)=0.4$. Note that, although there are noisy fluctuations in the exponent in the crossover regimes, the bulk and boundary exponents clearly manifest themselves as constants in Fig.~\ref{fig1}.
	
\begin{figure}
		\begin{minipage}{0.02\textwidth}
			\rotatebox{90}{\hspace{0.4in}$d\ln G/d\ln(V_0)$}
		\end{minipage}
		\begin{minipage}{0.45\textwidth}
			\centering
			\includegraphics[width=0.95\textwidth]{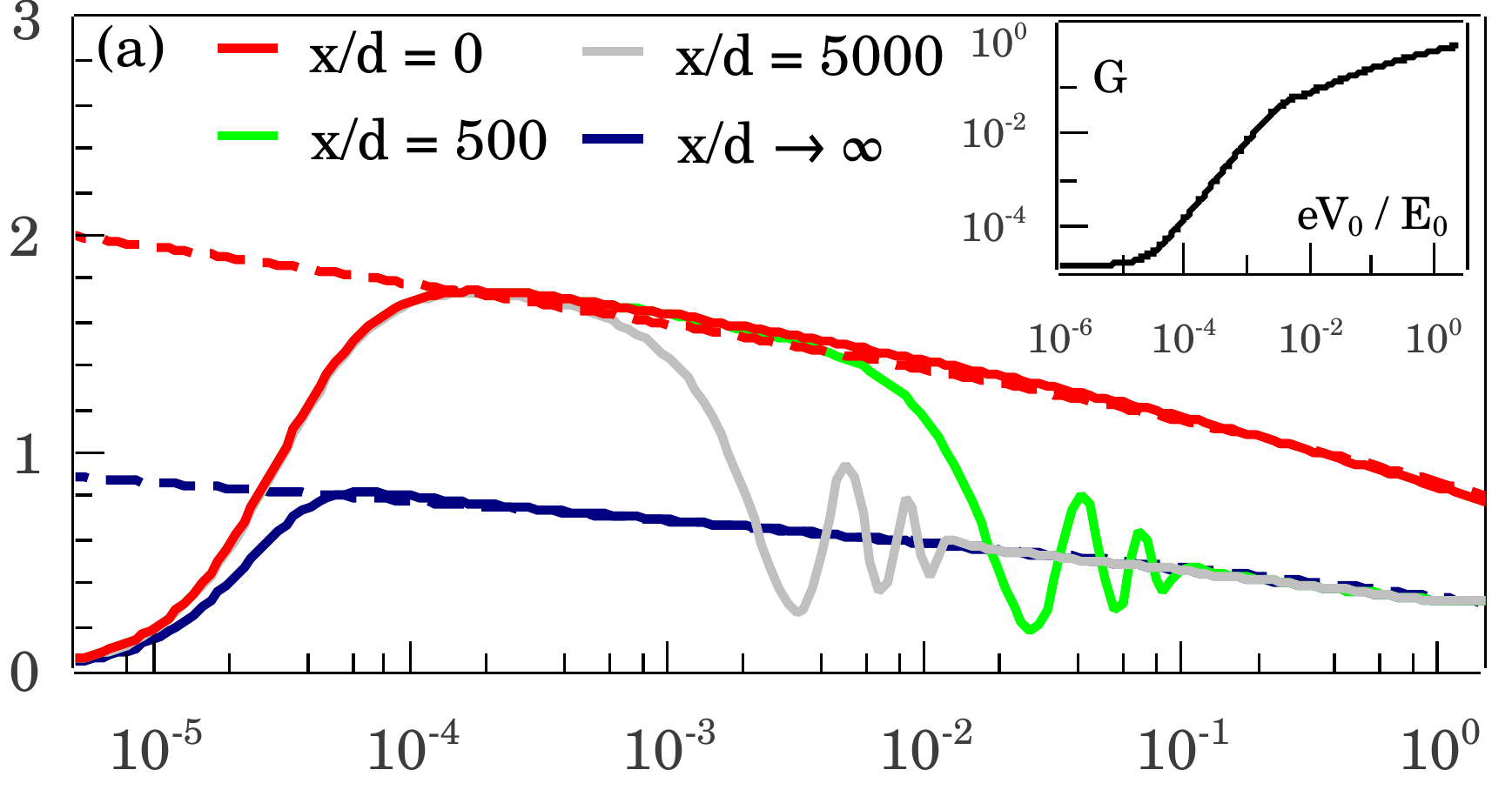}
			$eV_0/E_0$  	
			\vspace{0.1in} 		
		\end{minipage}
		\begin{minipage}{0.02\textwidth}
			\rotatebox{90}{\hspace{0.2in}$d\ln G/d\ln(T)$}
		\end{minipage}
		\begin{minipage}{0.45\textwidth}
			\centering
			\includegraphics[width=0.95\textwidth]{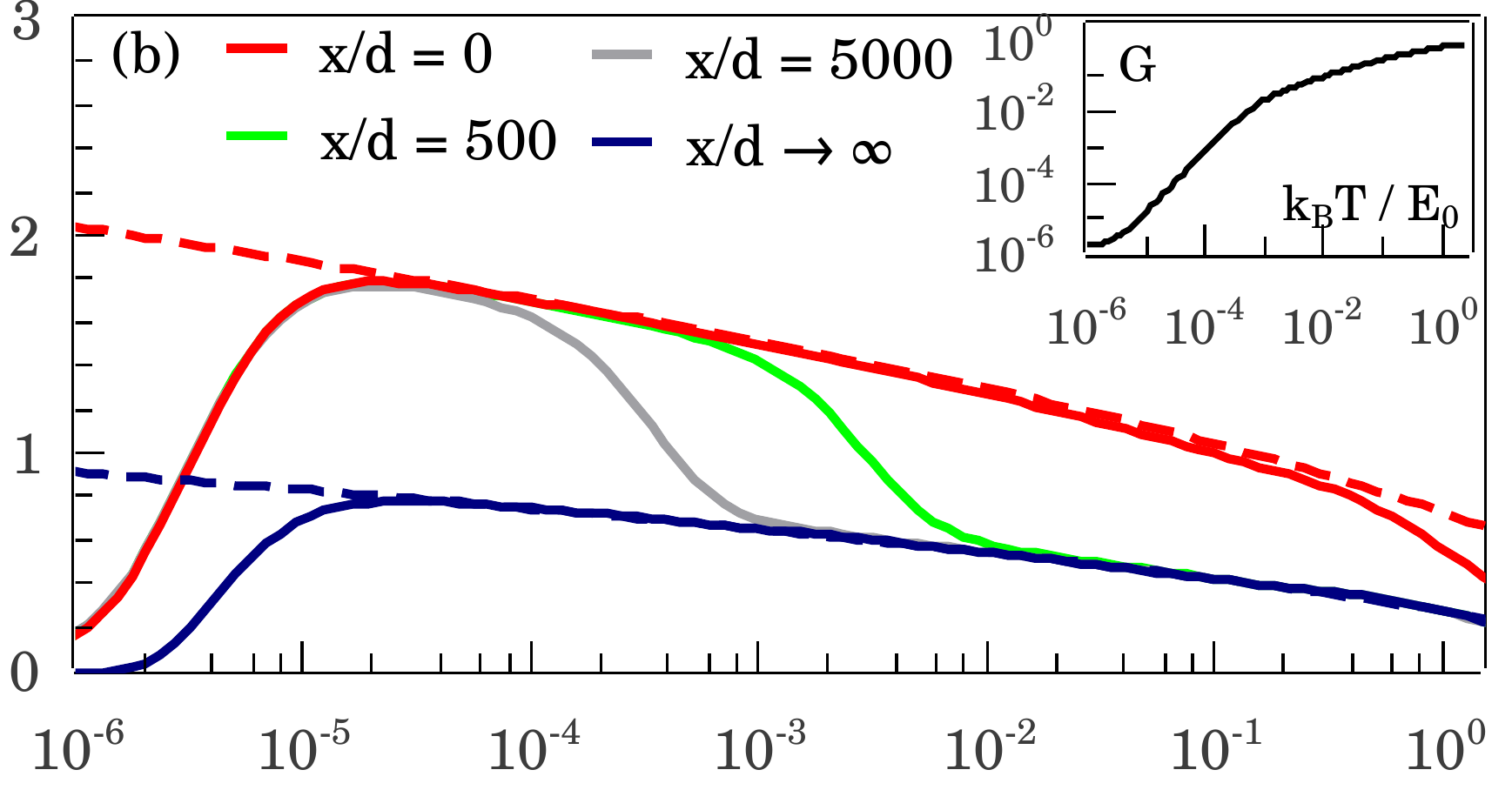}
			$k_BT/E_0$  	 		
		\end{minipage}			
		\caption{The effective exponent of the L-M tunneling conductance for a Coulomb Luttinger liquid with $U_0=5$ (a) with respect to $eV_0/E_0$ at fixed temperature $k_BT=10^{-5}E_0$ and (b) with respect to $k_BT/E_0$ at fixed voltage bias $eV_0=10^{-5}E_0$. The energy scale is $E_0=v_F/d$. The dashed lines are the theoretical effective exponent and the insets show the tunneling conductance for $x/d=2000$ in arbitrary unit.}\label{fig2}
	\end{figure}

    Because the scale-dependent Luttinger parameter in the Coulomb Luttinger liquid is an explicit function of the momentum, we can naturally define an energy scale $E_0=v_F/d$. In Fig.~\ref{fig2}, we show the directly calculated L-M effective exponent of a Coulomb Luttinger liquid with $U_0=5$. Similar to the short-range case, there are also three distinct regimes. However, the striking difference is the continuously increasing effective exponent with decreasing energy. For the conductance measurement with respect to the voltage bias $eV_0$ (see Fig.~\ref{fig2}(a)), the numerical result is consistent with Eq.~\eqref{eq24}. In addition, when $eV_0 \ll k_BT$ (see Fig.~\ref{fig2}(b)), the effective exponent of $G$ with respect to $T$ has the same form as Eq.~\eqref{eq24} with $\omega$ substituted by $k_BT$ and $\omega_s$ replaced by $T_s=7v_Fq_s\sqrt{U_0}$.
    
	\subsection{Universal scaling function}
\begin{figure}
		\centering
		\begin{minipage}{0.01\textwidth}
			\rotatebox{90}{\hspace{0.2in} Scaled $G$}
		\end{minipage}
			\begin{minipage}{0.45\textwidth}
				\centering
				\includegraphics[width=\textwidth]{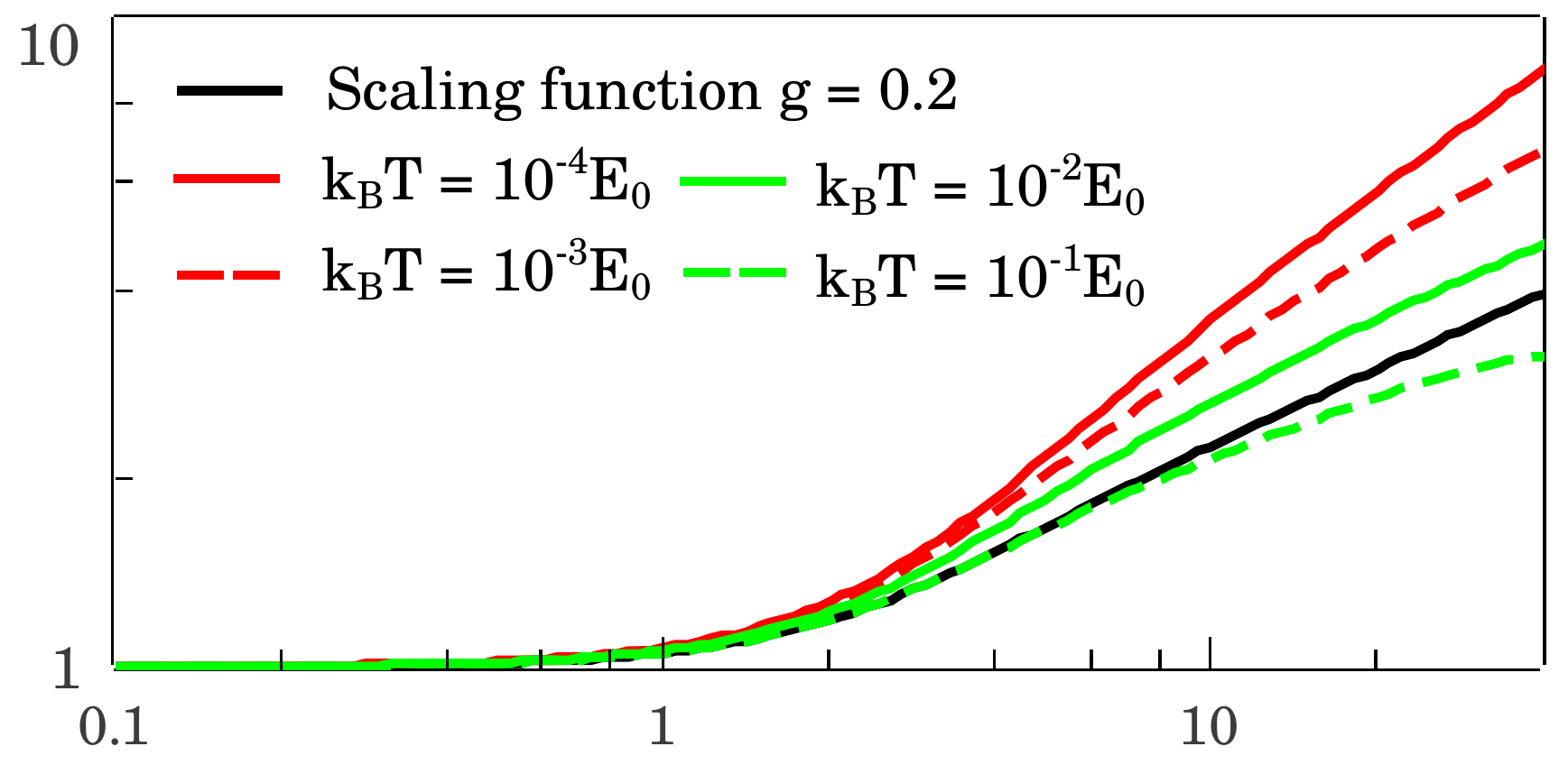}
				$eV_0/k_BT$  	 		
			\end{minipage}
			\caption{The scaled L-M tunneling conductances of a Coulomb Luttinger liquid with $U_0=5$ at different temperatures.}\label{fig3}	    	  	 	  
	\end{figure}

	Another contrasting property between the Coulomb and the short-range Luttinger liquid is the universal scaling with $eV_0/k_BT$, i.e. the tunneling conductances of a short-range Luttinger liquid at different temperatures can collapse into a single function of $eV_0/k_BT$. Using the asymptotic correlation function (no cut-off), the scaling function with respect to $\mu=eV_0/k_BT$ is \cite{Kane1992,Kane1992a,Postma2000}
	\begin{equation}
	\begin{split}
	&G_{L-L} \propto T^{\alpha}\sinh(\mu/2)\left| \Gamma(1+\alpha/2+i\mu/2\pi) \right|^2\\
	& \times \left\{\coth(\mu/2)/2 -\text{Im}\left[ \Psi(1+\alpha/2+i\mu/2\pi) \right]/\pi  \right\},\\
	&G_{L-M} \propto T^{\alpha}\cosh(\mu/2)\left| \Gamma((1+\alpha)/2+i\mu/2\pi) \right|^2;
	\end{split}
	\end{equation}
	where $\Gamma$ and $\Psi$ are the gamma and digamma functions. A single scaling function is a direct result of a scale invariant interaction constant $g$ where a conformal transformation can transform between the voltage and the temperature (corresponding to the real time and imaginary time boundary respectively). Clearly, the existence of a scale-dependent exponent for the Coulomb Luttinger liquid rules out such a universal scaling function uniquely determined by $eV_0/k_BT$. In Fig.~\ref{fig3}, we plot the scaled L-M tunneling conductance of a Coulomb Luttinger liquid at different temperatures, which clearly does not converge to a single function as in the short-range case. The predictions of Fig.~\ref{fig3} should also be directly experimentally verifiable using tunneling data provided that the varying range of the temperature and the bias voltage is of several orders of magnitudes.
	
\begin{figure}
		\begin{minipage}{0.01\textwidth}
			\rotatebox{90}{\hspace{0.2in} Scaled $G$}
		\end{minipage}
		\begin{minipage}{0.45\textwidth}
			\centering
			\includegraphics[width=\textwidth]{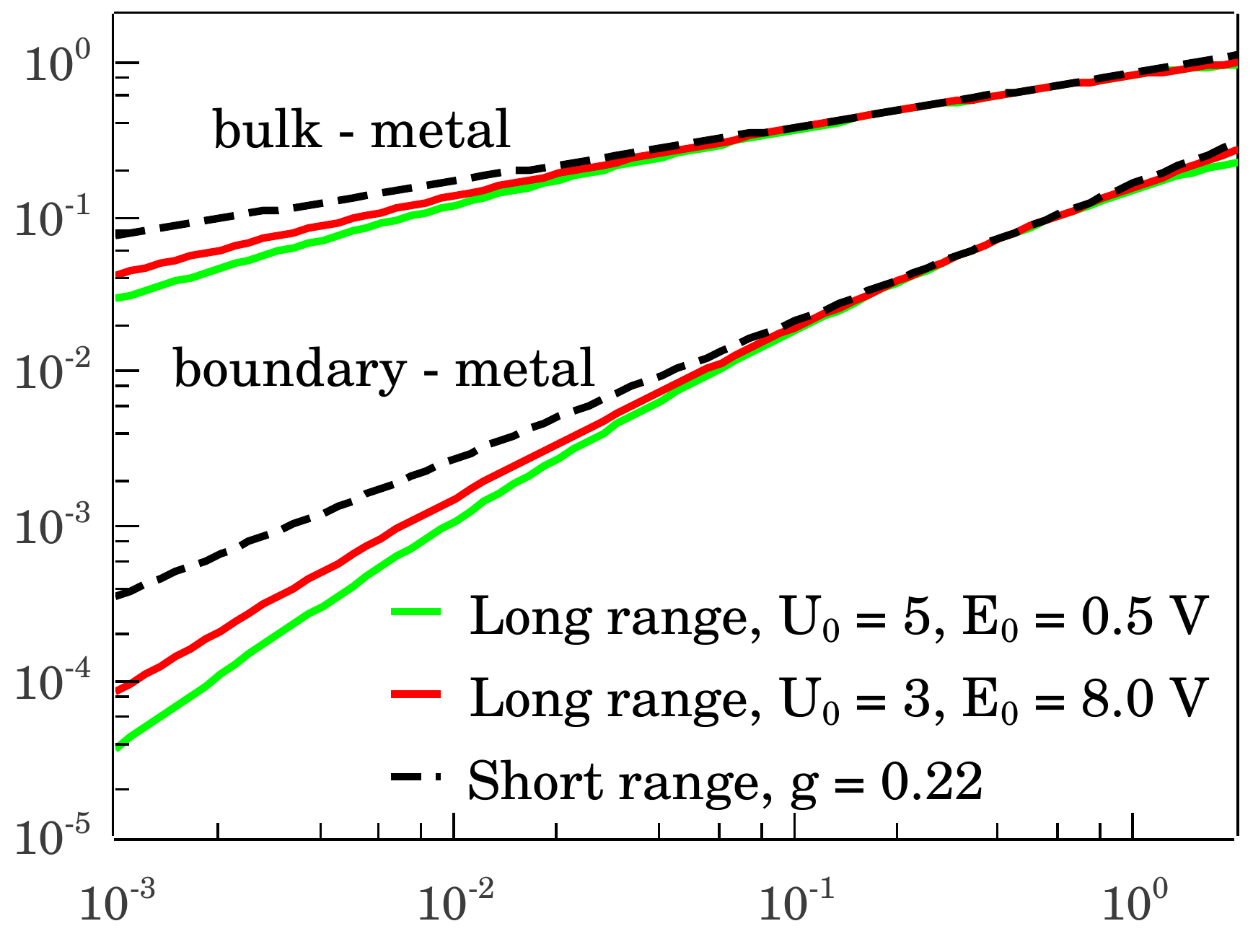}
			
			Voltage bias (V)  	 		
		\end{minipage}
		\caption{The scaled L-M tunneling conductances of a short-range  and two Coulomb Luttinger liquids (with different interaction parameters) at the bulk and at the boundary. Within one order of magnitude variation of the independent variable, the distinction among the three is hardly noticeable.}\label{fig4}	    	  	 	  
	\end{figure}

	We conclude this section by comparing the tunneling conductance of a short-range Luttinger liquid with two Coulomb Luttinger liquids characterized by different parameters at $T=0$. As can be seen from Fig.~\ref{fig4}, over a range of three orders of magnitude, the difference between short-range and long-range cases is obvious: in the log-log scale the tunneling conductance of a short-range interacting system is a line corresponding to an ideal power law while that for a Coulomb system is a downward bending curve characterizing a scale-dependent exponent. However, within a narrow range of one order of magnitude, the two are almost indistinguishable. Moreover, Coulomb Luttinger liquids  of different parameter sets are also indistinguishable within one order of magnitude of the tuning parameter. This suggests that we can only conclusively study the Coulomb Luttinger liquid if the dynamic range of the independent variable (i.e. $T, V_0$) is more than 2 orders of magnitude. We believe that the existing experimental literature on tunneling measurements is limited to a very narrow voltage and temperature range  and is thus insufficient to fully characterize the long-range interacting nature of Coulombic systems.  A true verification of the theory necessitates the observation of the scale dependence shown in Figs.~\ref{fig2},~\ref{fig3}, and \ref{fig4}, which would require a large variation in the dynamical range of temperature and bias voltage. We suggest tunneling experiments be carried out with a substantial increase in the dynamical range of temperature and voltage so that the predicted scale-dependent deviation from a constant Luttinger exponent can manifest itself in the measurement.
	
	\subsection{High-energy crossover to free Fermi gas}
\begin{figure}
		\begin{minipage}{0.01\textwidth}
			\rotatebox{90}{\hspace{0.2in} Scaled $G$}
		\end{minipage}
		\begin{minipage}{0.23\textwidth}
			\centering
			\includegraphics[width=\textwidth, height=\textwidth]{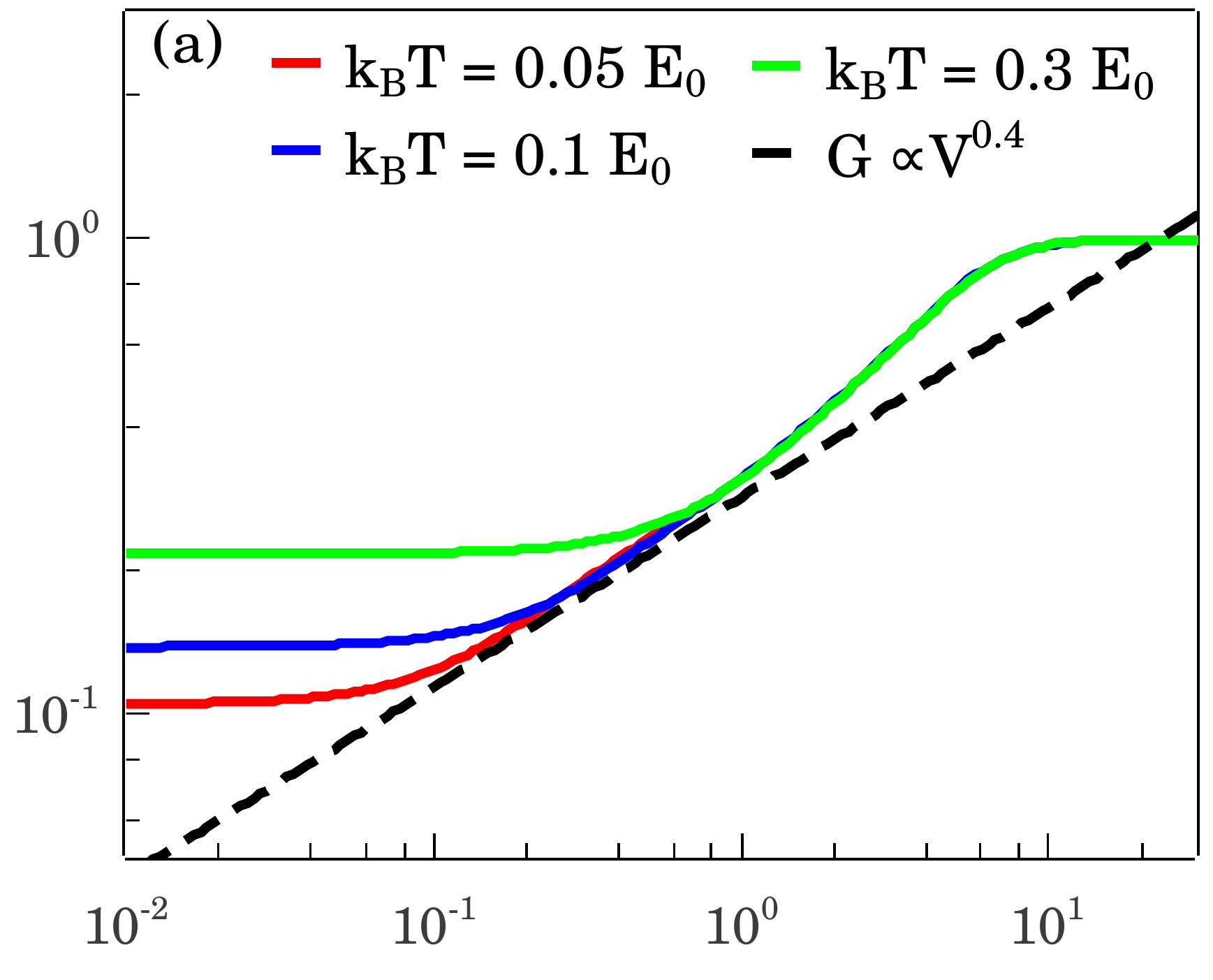}
			\includegraphics[width=\textwidth, height=\textwidth]{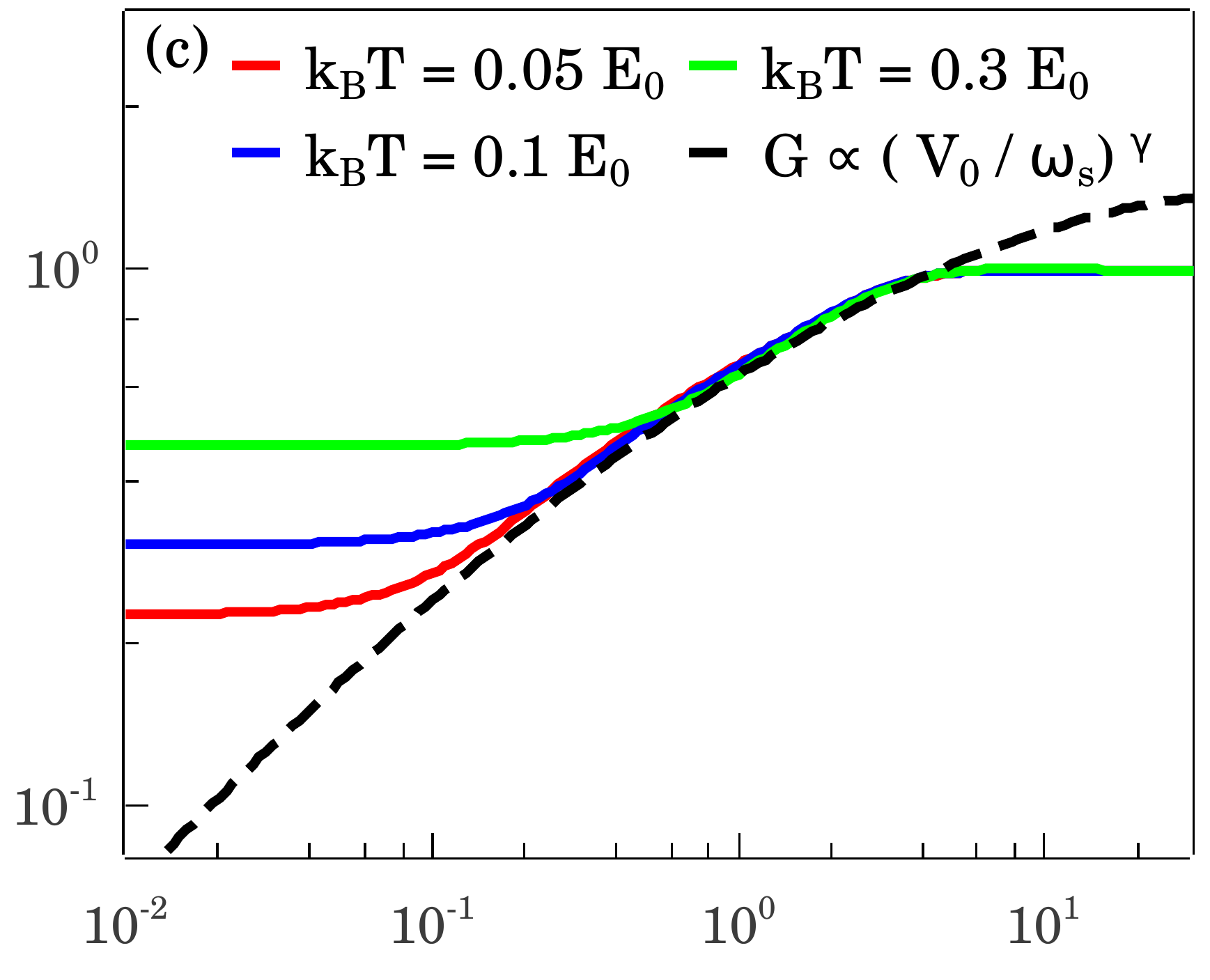}
			$eV_0/E_0$							  	 		
		\end{minipage}
		\begin{minipage}{0.23\textwidth}
			\centering
			\includegraphics[width=\textwidth, height=\textwidth]{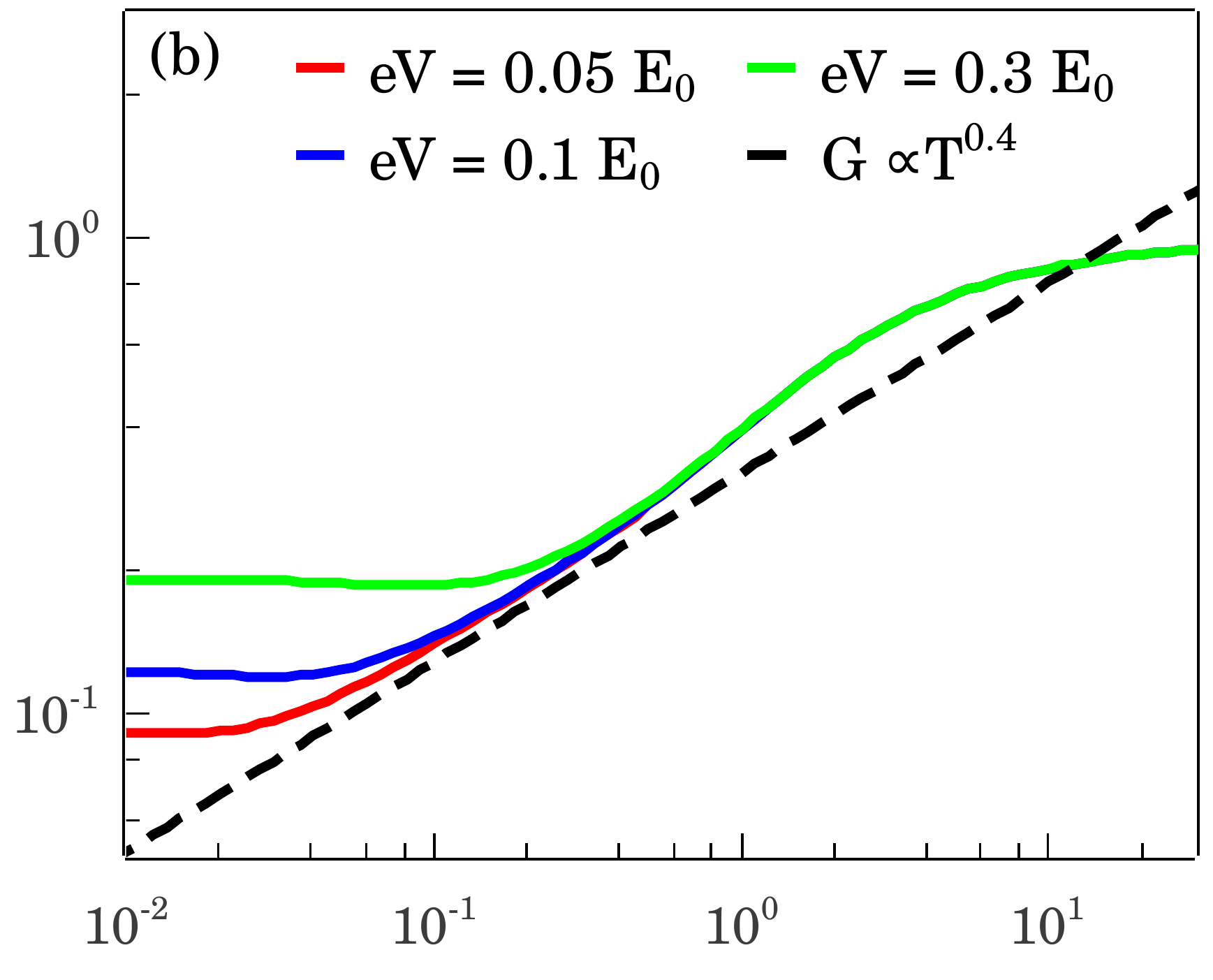}
			\includegraphics[width=\textwidth, height=\textwidth]{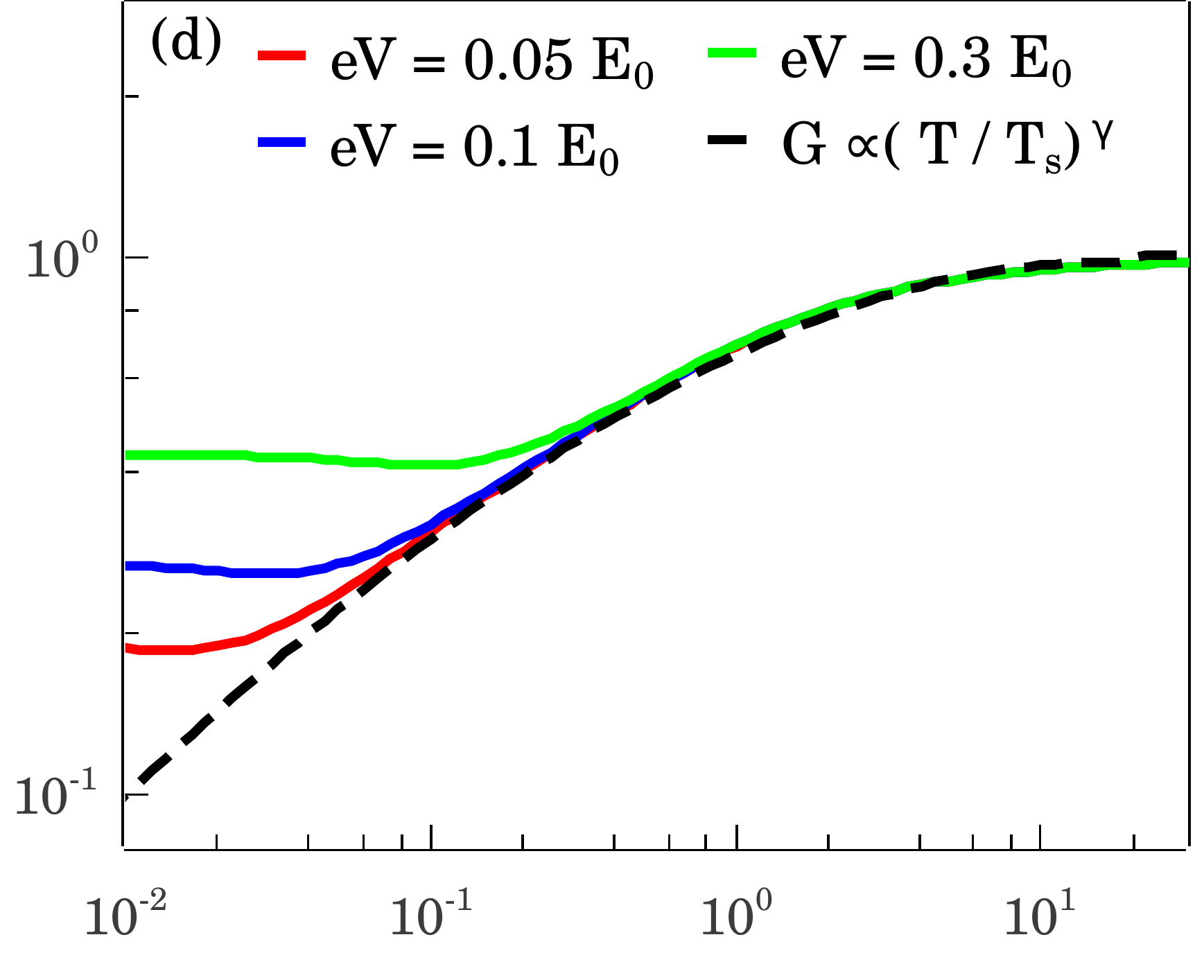}
			$k_BT/E_0$							  	 		
		\end{minipage}		
		\caption{The scaled L-M bulk tunneling conductances of a short-range Luttinger liquid with $g=0.2$ as a function of $eV_0$ (a) and $k_BT$ (b). The scaled L-M bulk tunneling conductances of a Coulomb Luttinger liquid with $U_0=5$ as a function of $eV_0$ (c) and $k_BT$ (d). The dashed lines are theoretical predictions without regarding the high energy crossover.}\label{fig5}	  		
	\end{figure}

	If we assume that the interaction vanishes at very high momentum, which is true for all interactions with ultraviolet regularization, then the Luttinger liquid can cross over to the free Fermi gas defined by $g\to 1$. In this section, we investigate the system near this high-energy crossover. For the short-range model defined previously, it is obvious that the crossover energy must be associated with the cut-off momentum $\Lambda$ while for the Coulomb interaction, the interaction potential $K_0(qd)$ decays exponentially for $q\gg 1/d$ so the crossover energy must be related to $1/d$. Therefore, for the Coulomb Luttinger liquid, the crossover energy scale is physically defined whereas for the short-range model, it depends on an ad hoc cut off. Hence, the energy scale we define earlier, i.e. $E_0=v_F\Lambda$ for short-range interaction and $E_0=v_F/d$ for Coulomb interaction, is also the high-energy crossover scale. In Fig.~\ref{fig5}, we show the L-M tunneling conductance as a function of the voltage bias and temperature. The usable region where one can extract a meaningful power law is for $eV_0 (k_BT)$ much higher than $k_BT (eV_0)$ but still less than the crossover scale $E_0$. As a result, if $eV_0$ is close to $E_0$, one cannot extract the power law with respect to the temperature and vice versa. Moreover, in the case of Coulomb Luttinger liquid, the theoretical scale-dependent exponent already intrinsically contains a decay at high energy even though it is derived using low energy assumption. This is because the definition of the long-range interaction already contains the ultra-violet regularization through the transverse size $d$. Therefore, the consistency between numerical results and theoretical predictions extends to much higher energy than the short-range interacting Luttinger liquid. The fact that the Coulomb Luttinger liquid has a built-in physical ultraviolet regularization, in contrast to a completely arbitrary cut-off-dependent regularization in the short-range model, makes the long-range interaction model more theoretically meaningful.
	
	\section{Finite-size effect}
	In this section, we study the effect of finite size (i.e. the finite length of the 1D system in the experimental tunneling measurements) on the experimentally extracted properties of the Luttinger liquid, e.g. the momentum distribution and the density of states. We will show that the finite system (as long as the system length is much longer than the ultraviolet cutoff length $d$) does not alter the Luttinger liquid properties except for inducing discrete peaks pattern to the measured spectrum. This effect, however, can only be observed if experiments have resolution much better than the level splitting, which is typically not the case. Since this discreteness has not been reported in experiments, the finite-size effect may not be relevant to experiments. Beside the finite resolution, finite temperature most likely smoothens the level spacing induced peak structures in the actual experiments.  For very short wires, the system would behave as a quantum dot dominated by Coulomb blockade and the Luttinger liquid behavior becomes irrelevant. For simplicity, we consider a 4-fold degenerate 1D system with size $L$ and periodic boundary condition. The use of the periodic boundary condition in this section (in contrast to the rest of this work) is to separate the finite-size effect from the non-trivial perturbation of the open boundary.
	
	\subsection{Momentum distribution function}
    We first consider the chiral static correlation function 
	\begin{equation}
	\begin{split}
 	 C(x,0) =\lim\limits_{\epsilon\to 0^+} \frac{e^{i\pi x/L}}{L\left[1-e^{(ix-\epsilon)q_0}\right]}e^{- \sum_{n=1}^\infty H(n)},
	\end{split}
	\end{equation}
	where
	\begin{equation}
		H(n)=\frac{1-\cos(qx)}{n}\frac{g(q)+q^{-1}(q)-2}{8},
	\end{equation}
	 $q_0=2\pi/L$ and $q=nq_0$. The momentum distribution at momentum $k$ with respect to $k_F$ is
	\begin{equation}
	\begin{split}
	  n(k)&=\int e^{-ikx} C(x,0)dx\\
	  	&=\frac{1}{2}-\frac{1}{2L}\int \frac{\sin kx}{\sin(q_0 x/2)}e^{-\sum_n H(n)}dx.
	\end{split}
	\end{equation} 
	To retrieve the infinite-size limit, one can take the limit $q_0\to 0$, in which $q_0/(2\sin(q_0x/2))\approx 1/x +\mathcal{O}(qx)^2$ and the sum in the exponential is replaced by the corresponding integral. The difference between the sum and the integral counterpart is given by the Euler-Maclaurin formula
	\begin{equation}
	\begin{split}
	  &\sum_{n=1}^{\infty}H(n) - \int_{0}^{\infty}H(n)dn \\
	  &= \int_{0}^{1}H(n)dn+ \frac{H(1)}{2}-\frac{H'(1)}{12} + \mathcal{O}(H''(1))\\
	  &= -\frac{q_0^2x^2\alpha(q_0)}{24} + \mathcal{O}(q_0x)^3
	\end{split}
	\end{equation}
	with $\alpha(q_0)=(g(q_0)+g^{-1}(q_0)-2)/8$. We emphasize that even for the logarithmically divergent $\alpha(q_0)\sim \ln^{1/2}(q_s/q_0)$ in the Coulomb Luttinger liquid, this divergence is much weaker than the quadratic decay $q_0^2$. As a result, the finite-size correction to the momentum density function has the order of $\mathcal{O}(1/kL)^2$. 
	
\begin{figure}
		\begin{minipage}{0.02\textwidth}
			\rotatebox{90}{\hspace{0.2in} $n_{exp}(k)-1/2$}
		\end{minipage}
		\begin{minipage}{0.45\textwidth}
			\centering
			\includegraphics[width=0.95\textwidth]{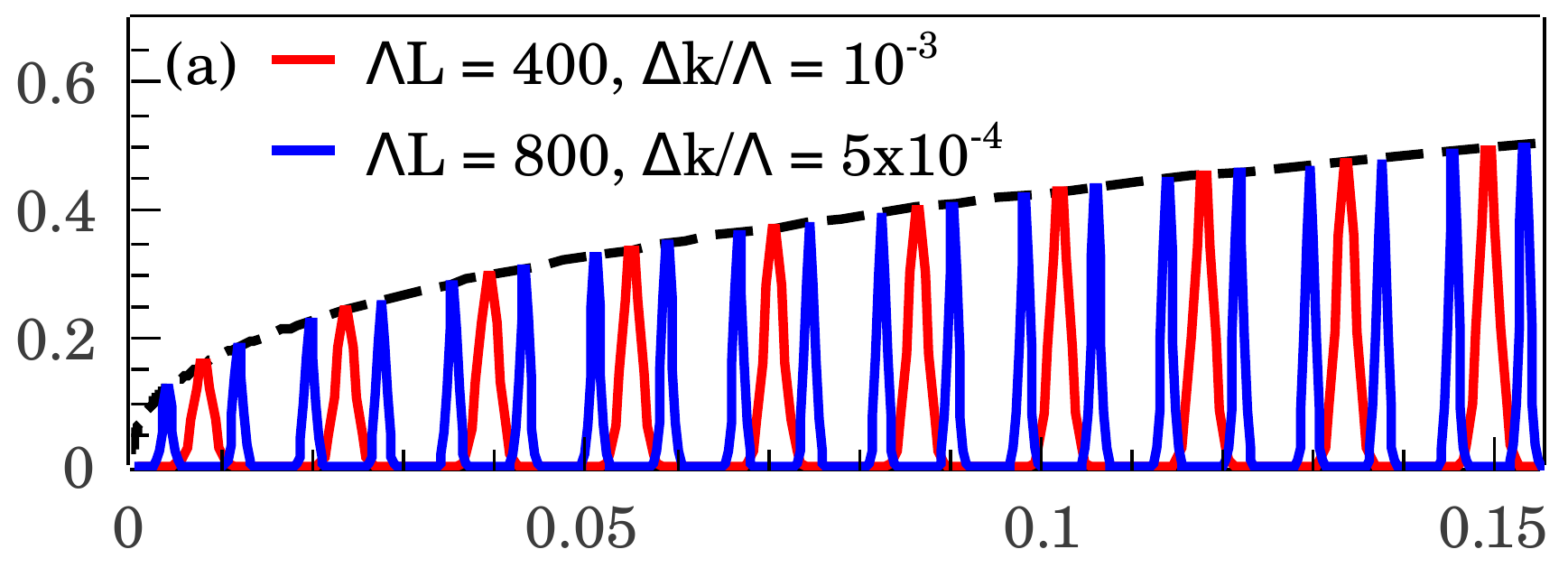}
			\includegraphics[width=0.95\textwidth]{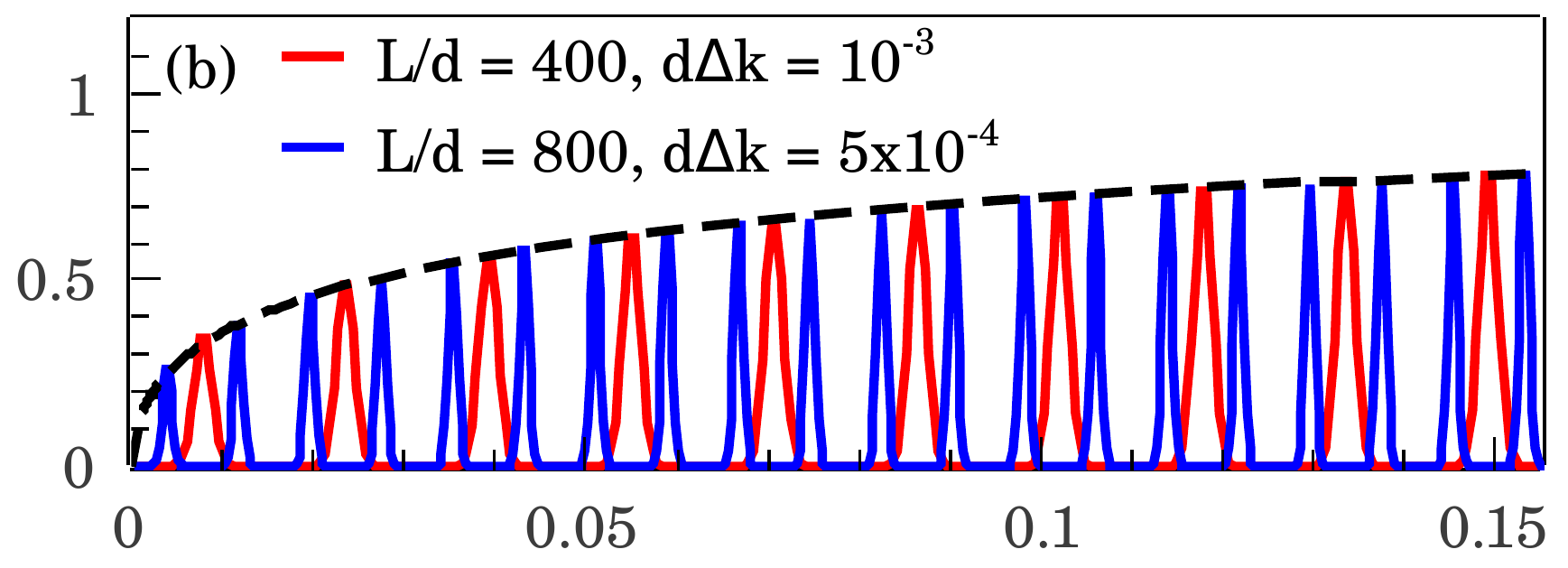}
			$k/k_0$ 			 	 		
		\end{minipage}
		\caption{The normalized momentum distribution of (a) short-range interacting Luttinger liquid with $g=0.2$ and (b) Coulomb Luttinger liquid with $U_0=5$. The momentum scale $k_0=\Lambda$ for the short-range and $k_0=1/d$ for the Coulomb Luttinger liquid. The dashed lines are the corresponding momentum distribution in the limit $L\to \infty$.}\label{fig7}	    	  	 	  
	\end{figure}

   In practice, measurements have finite resolution (and finite temperature) that smooths out the physical quantity and removes fine details. The effect of the finite resolution can be presented by a convolution
	\begin{equation}
	\begin{split}
	  n_{exp}(k) &= \int n(k-k')S(k')dk'\\
	   &= \int e^{-ikx}C(x,0)\tilde{S}(x)dx
	   \end{split}\label{eq28}
	\end{equation}
	where $S(k)$ is a distribution function and $\tilde{S}(x)$ is the Fourier transform of $S(k)$. If we assume $S(k)$ is a Gaussian function with standard deviation $\Delta k$, then $\tilde{S}(x)=e^{-\Delta k^2x^2/2}$. If $\Delta k L \ll 1$, there are several periods of $C$ inside the integral interval, leading to the peak pattern in $n_{exp}(k)$ at $k=(n+1/2)q_0$ with $n$ being an integer.  We show in Fig.~\ref{fig7} the momentum distribution of finite-length short-range and Coulomb Luttinger liquids compared with their infinite-length counterparts. At the resonant momentum $kL=(2n+1)\pi$, the finite-size correction $ \sim 1/((2n+1)\pi)^2$ is insignificant. Therefore, the peaks at resonant momentum match the momentum distribution of the corresponding infinite system as shown in Fig.~\ref{fig7}.
	
\begin{figure}
		\begin{minipage}{0.02\textwidth}
			\rotatebox{90}{\hspace{0.2in} $d\ln(n_{exp}(k)-1/2)/d\ln(k)$}
		\end{minipage}
		\begin{minipage}{0.45\textwidth}
			\centering
			\includegraphics[width=0.95\textwidth]{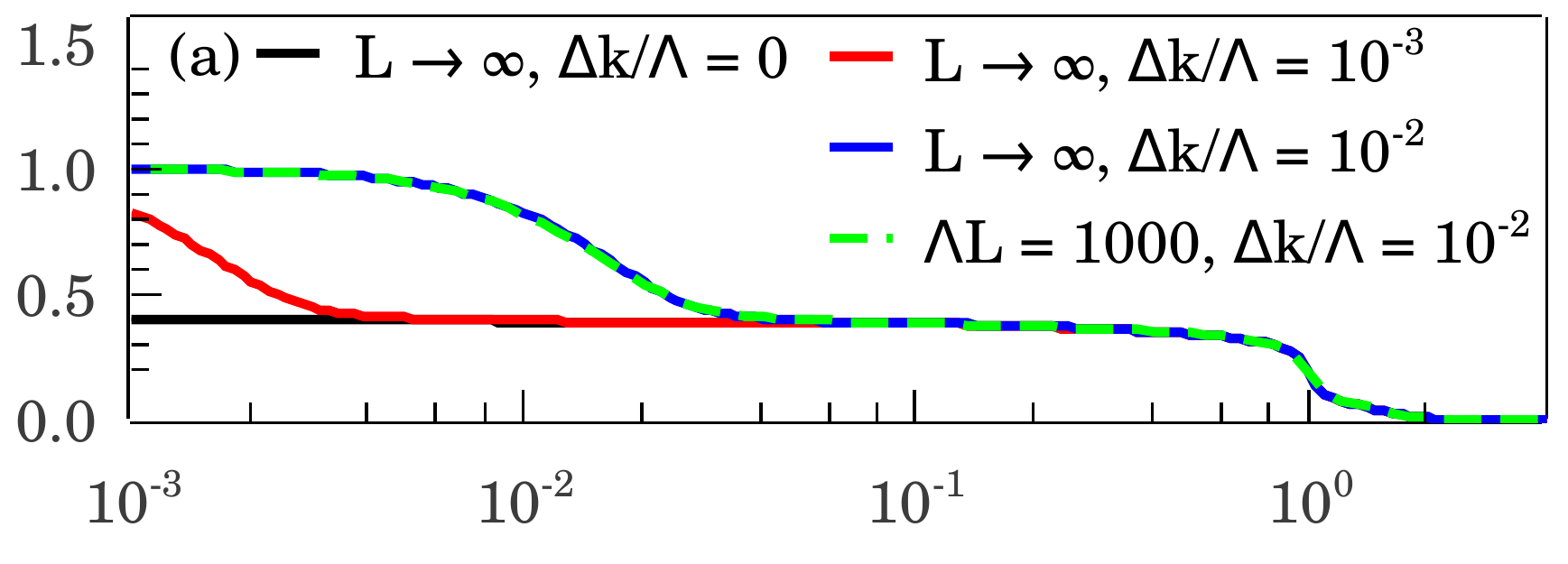}
			\includegraphics[width=0.95\textwidth]{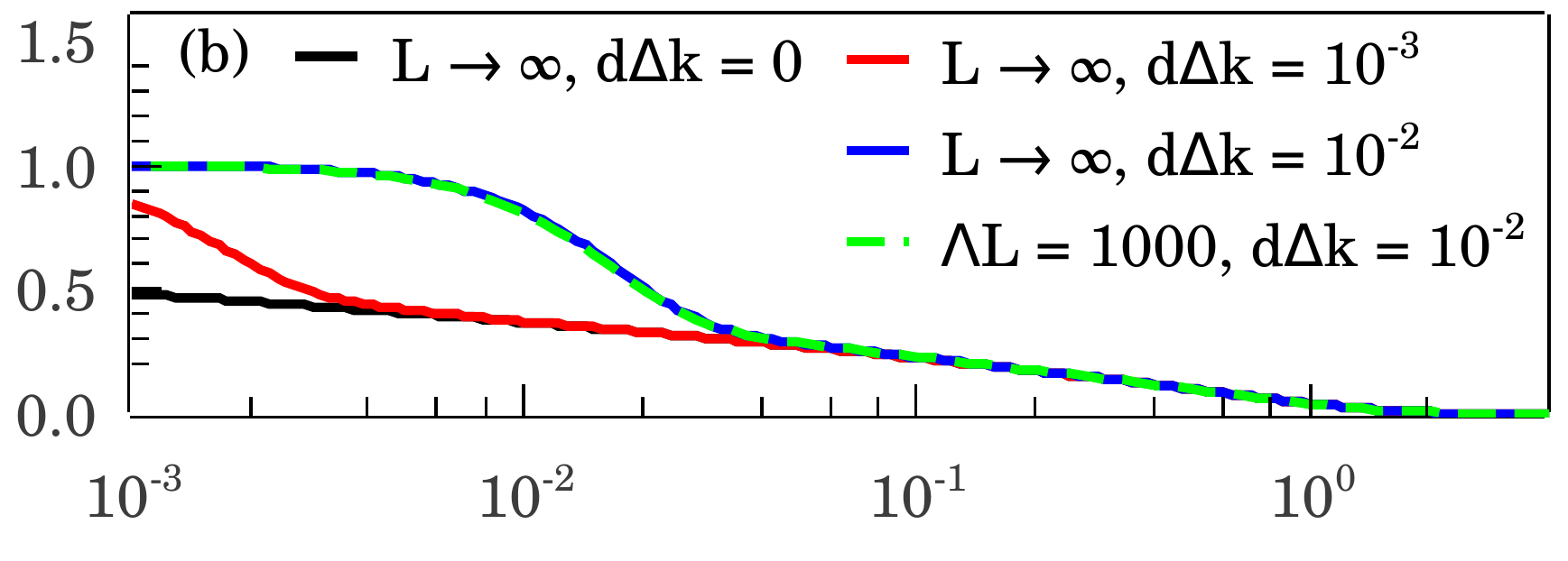}
			$k/k_0$ 			 	 		
		\end{minipage}
		\caption{The effective exponent of the momentum distribution for (a) short-range interacting Luttinger liquid with $g=0.2$ and (b) Coulomb Luttinger liquid with $U_0=5$. This exponent always approaches 1 for $k < \Delta k$. The scale $k_0$ is similarly defined as in Fig.~\ref{fig7}.}\label{fig8}	    	  	 	  
	\end{figure}

	Next, we consider the second case when $\Delta k L \ll 1$ and the measured spectrum is smooth. Then, we can calculate the effective exponent by taking the first derivative $d\ln(n_{exp}(k)-1/2)/d\ln(k)$ and display the results in Fig.~\ref{fig8}. An interesting feature is how the finite resolution affects even the distribution in infinite-length cases. It is known that $n(k)-1/2 \sim k^\alpha$ has a singularity in the first-order derivative. The convolution, regardless of the exact form of the distribution function, always suppresses this singularity, thus making $n_{exp}(k)-1/2 \sim k$ and the effective exponent approaches 1 for $k < \Delta k$. For the finite-size case, the non-trivial behavior only manifests for $k>\Delta k$ but the finite-size correction is of the order $1/(k L)^2 < 1/(\Delta k L)^2\ll 1$, rendering its effective exponent almost identical to that of the corresponding infinite-size case. 

	\subsection{Finite-size density of states}
\begin{figure}
	\begin{minipage}{0.02\textwidth}
		\rotatebox{90}{\hspace{0.2in} $\rho_{exp}(\omega)v_F$}
	\end{minipage}
	\begin{minipage}{0.45\textwidth}
		\centering
		\includegraphics[width=0.95\textwidth]{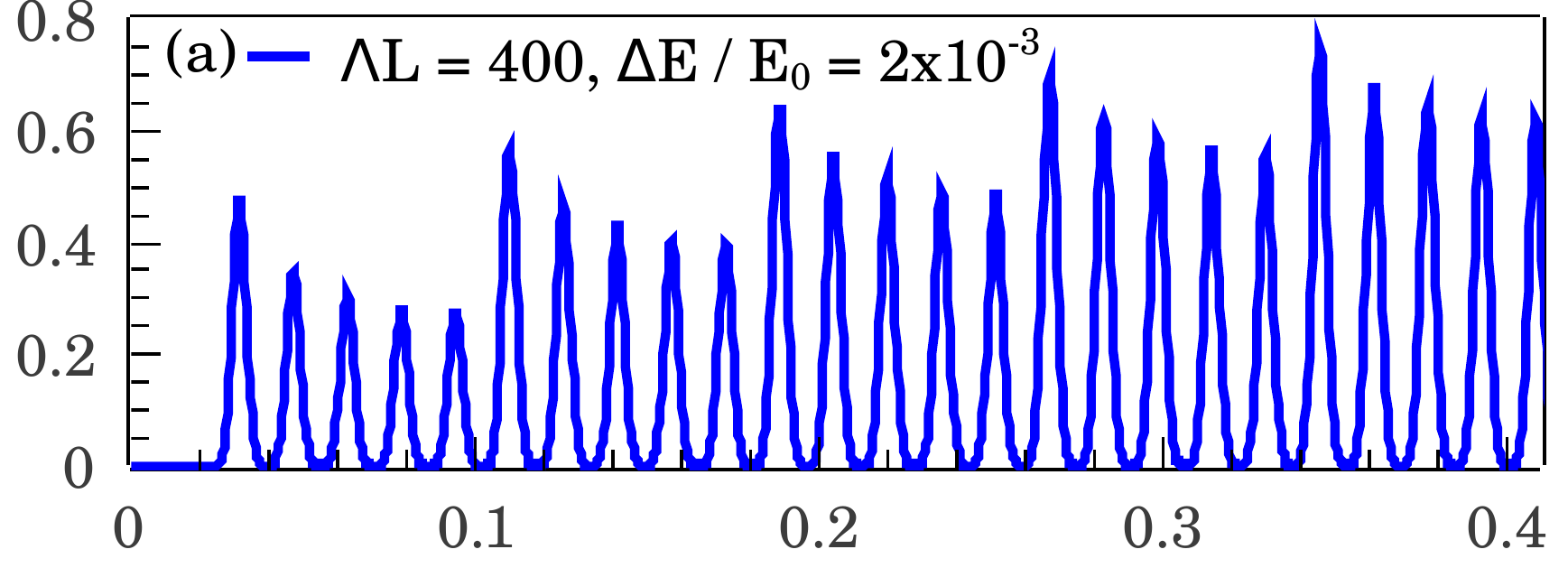}
		\includegraphics[width=0.95\textwidth]{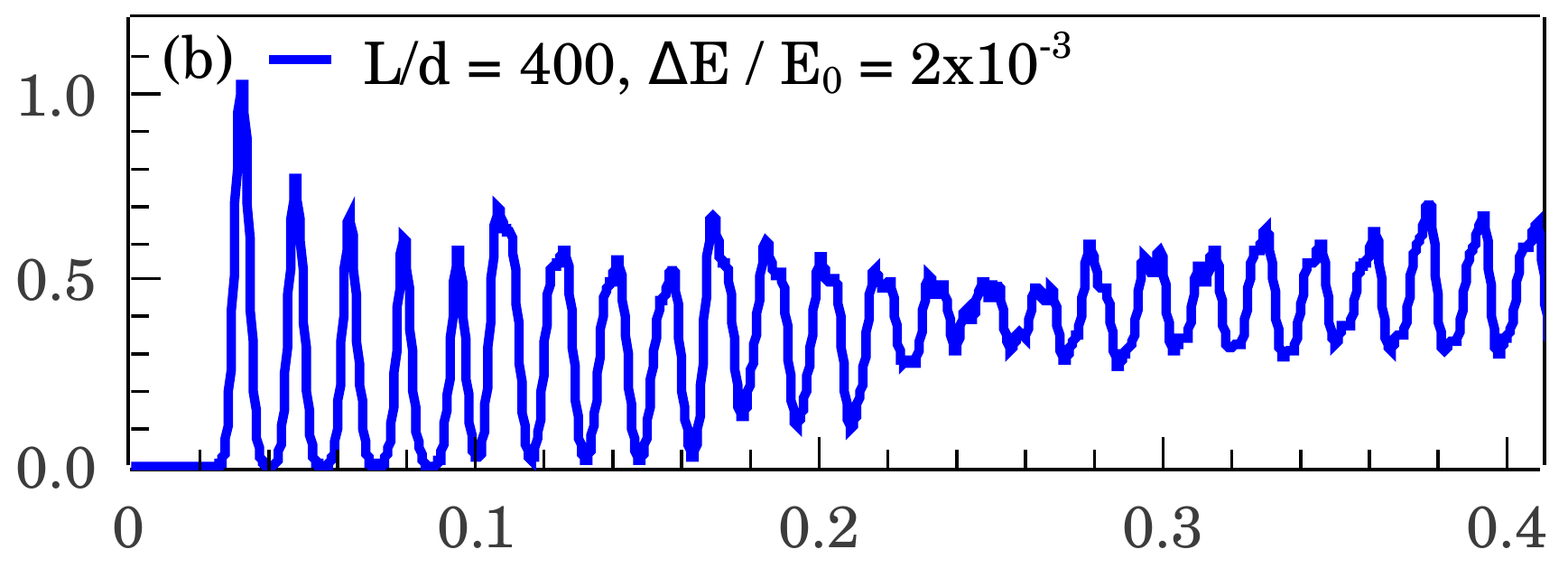}
		$\omega / E_0$ 			 	 		
	\end{minipage}
	\caption{The density of states of (a) short-range interacting Luttinger liquid with $g=0.2$ and (b) Coulomb Luttinger liquid with $U_0=5$. The interference pattern is caused by different existing excitation propagating velocities.}\label{fig9}	    	  	 	  
\end{figure}

 We now study the finite-size effect on the density of states, which has direct relation to tunneling experiments. Analogous to the momentum distribution, we mimic the finite resolution effect by introducing a decaying function $e^{-\Delta E^2 t^2/2}$ into the  dynamic correlation function
\begin{equation}
\begin{split}
& C(0,t) =\lim\limits_{\epsilon\to 0^+} \frac{e^{-i(3E_u+E_\rho)t-\Delta E^2 t^2/2}}{L\left[1-e^{(-iv_Ft-\epsilon)q_0}\right]}e^{\sum_{n=1}^{\infty} -J(n)},
\end{split}
\end{equation}
where 
\begin{equation}
	J(n) =  \frac{1-e^{-iv_\rho q t}}{n} \frac{g(q)+g^{-1}(q)}{8} -\frac{1-e^{-iv_Fqt}}{4n},
\end{equation}
and the charge gaps
\begin{equation}
E_u=\frac{1}{4}\frac{q_0v_F}{2},\quad  E_\rho= \frac{1}{4}\frac{q_0v_F}{2}\frac{1+g(q_0)^{-2}}{2}.
\end{equation}
The first term corresponds to three unrenormalized spin/valley channels, while the second one represent the renormalized charge excitation. The density of states as a function of $\omega=E-E_F$ is defined through the Fourier transform
\begin{equation}
\begin{split}
 &\rho_{exp}(\omega) = \frac{1}{2\pi} \int e^{i\omega t}\left[C(0,t) + C(0,-t)\right]dt.
\end{split}
\end{equation}  
For the short-range interacting Luttinger liquid, we can easily see that the $C(0,t)$ has two periods: $t_1=L/v_F$ for the three un-renormalized channels and the high-energy plasmon channel when $\omega \gg v_F\Lambda$, and $t_2=L/v_\rho$ for the renormalized plasmon channel. As a result, the Fourier transform for $L\Delta E/v  \ll 1$ shows two groups of peaks in Fig.~\ref{fig9}a. The group of major peaks has the gap of $q_0v_\rho$, corresponding to the plasmon excitation; the second one of minor peaks has the gap of $q_0v_F$ reflecting excitations in the other channels. Due to this interference, it is not possible to compare the finite-size $\rho_{exp}(\omega)$ with its infinite-size counterpart. In fact, if we revert to the spinless model, i.e. removing other excitation channels, and push $\Lambda \to \infty$ one can fit the peak pattern of the finite-$L$ into $\rho(\omega)$ computed in the infinite-size model. For the Coulomb Luttinger liquid as in Fig.~\ref{fig9}b, the plasmon velocity is already scale-dependent, as a result, the density of state shows intricate interference pattern.

\begin{figure}
	\begin{minipage}{0.02\textwidth}
		\rotatebox{90}{\hspace{0.2in} $d\ln(\rho_{exp}(\omega))/d\ln(\omega)$}
	\end{minipage}
	\begin{minipage}{0.45\textwidth}
		\centering
		\includegraphics[width=0.95\textwidth]{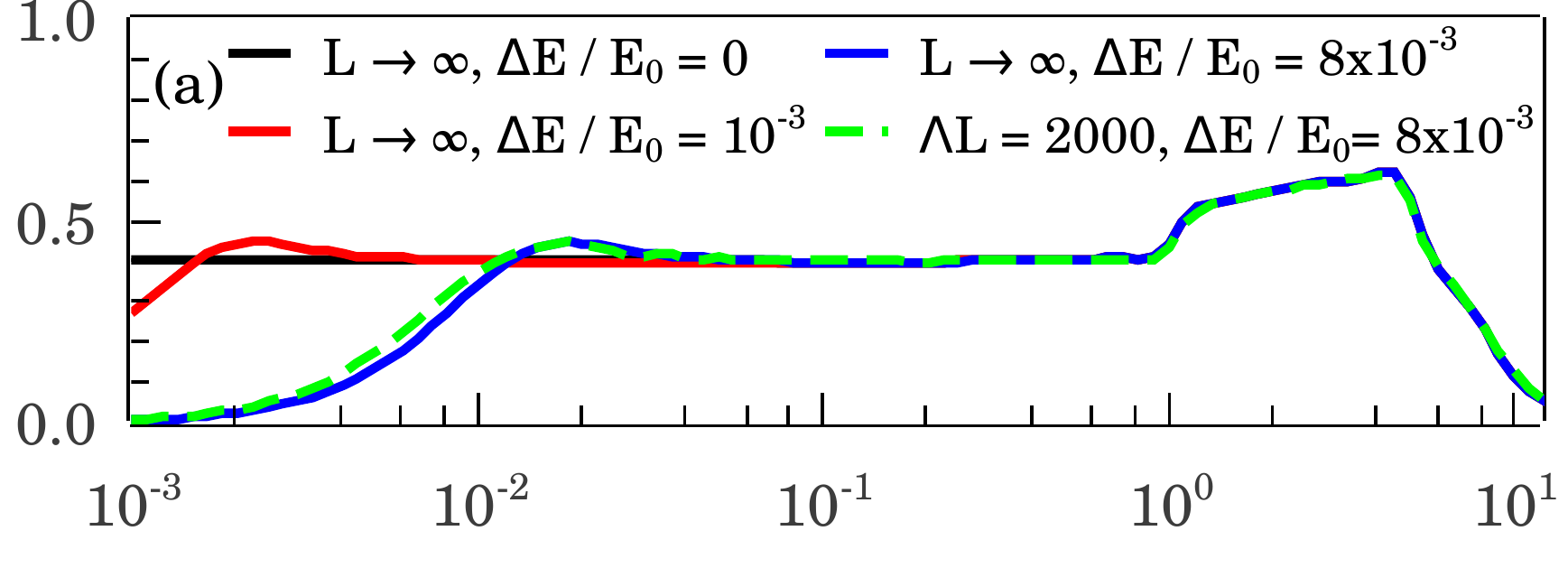}
		\includegraphics[width=0.95\textwidth]{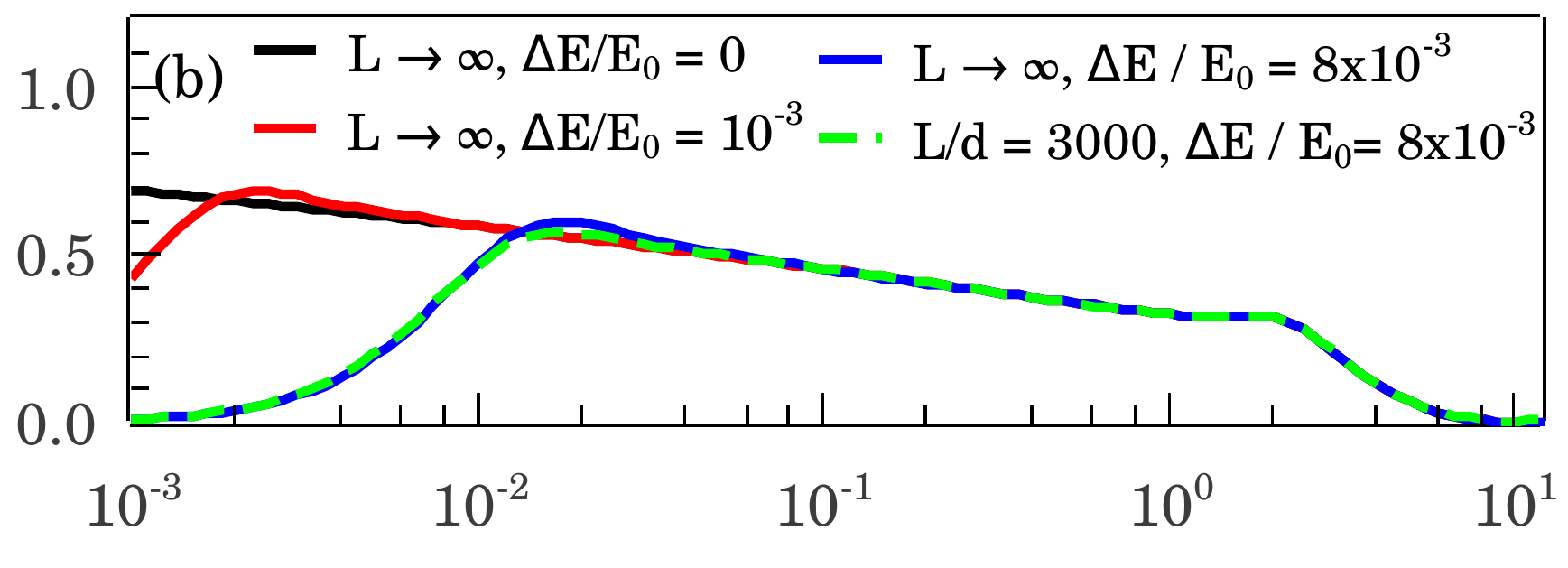}
		$\omega/E_0$ 			 	 		
	\end{minipage}
	\caption{The effective exponent of the density of states for (a) short-range interacting Luttinger liquid with $g=0.2$ and (b) Coulomb Luttinger liquid with $U_0=5$. This exponent vanishes for $\omega < \Delta E$.}\label{fig10}	    	  	 	  
\end{figure}

We now study the smooth regime for $L \Delta E/v  \gg 1$. The density of states is given by
\begin{equation}
\begin{split}
 	\rho_{exp}(\omega)v_F &= 1-\frac{v_F}{L}\int \frac{e^{-\Delta E^2 t^2/2} \cos\omega t }{\sin(q_0v_Ft/2)}\\
 	&\times \text{Im}\left(e^{-\sum J(n)-i(E_\rho-E_u)t}\right)dt.
\end{split}\label{eq35}
\end{equation}
Applying the same technique for the estimating the momentum distribution,
\begin{equation}
	\begin{split}
&\sum_{n=1}^{\infty}J(n) - \int_{0}^{\infty}J(n)dn \\
&= \int_{0}^{1}J(n)dn+ \frac{J(1)}{2}-\frac{J'(1)}{12} + \mathcal{O}(J''(1))\\
&= \frac{-iq_0t}{2}\left[\frac{v_\rho(q_0)(g(q_0)+g^{-1}(q_0))}{8} - \frac{v_F}{4} \right] \\
&- \frac{(q_0t)^2}{24}\left[\frac{v^2_\rho(q_0)(g(q_0)+g^{-1}(q_0))}{8} - \frac{v_F^2}{4} \right]+\mathcal{O}(q_0v_Ft)^3. 
\end{split}\label{eq36}
\end{equation}
The first-order term of Eq.~\eqref{eq36} exactly cancels the charge gap term $i(E_\rho-E_u)t$ in Eq.~\eqref{eq35}, thus the finite-size effect only introduce corrections of the order of $(v/\omega L)^2$ or higher into the density of states. As shown in Fig.~\ref{fig10}, even for infinite-size systems, the effective exponent approaches zero for $\omega < \Delta E$. This is because the convolution fills out the singular pseudo gap at $\omega=0$, making this gap effectively a non-zero constant. For finite-size systems, non-trivial behavior only appears for $\omega > \Delta E$, with a negligible correction proportional to $(v/\Delta E L )^2$. We note that for a finite-system with open boundaries, the same conclusion can be made; however, the transition from the bulk to boundary exponent happens with $\pi v(\sin(\pi x/L)L)^{-1}$ replacing $v/x$ as in the limit $L\to \infty$. In fact, the suppressed influence of the finite size on the observed density of states has been mentioned in the literature before \cite{kane, boundary1, boundary2}, but our results show quantitatively that the finite size effects are not a serious problem for 1D Lutinger liquid studies, either for the short-range or the long-range model.

	\section{Conclusion}
In this paper, we theoretically compare the tunneling conductance in 1D systems between the short-range and the long-range interacting Luttinger liquid models. The logarithmic divergence of the long-range Coulomb interaction gives raise to a scale-dependent effective exponent which increases at lower energy. However, this exponent varies slowly, giving an impression of an actual power law when the dynamic range (of temperature or bias voltage) is around one order of magnitude or less. The difference between short-range and long-range Luttinger liquid conductance is more visible over the range of two or more orders of magnitude in the independent variable.  We believe that the clear theoretical difference between short- and long-range Luttinger liquids established in this paper should be experimentally observable provided that the experimental tuning variables (temperature and bias voltage) are varied over a large dynamical range. 

We also study the effect of the high energy crossover to the free Fermi gas and the finite-size effect. This high-energy crossover is due to short-distance behavior of the interaction and depends on the microscopic ultraviolet regularization. Near the high-energy crossover, the observed tunneling conductance deviates from the low-energy theoretical prediction and approaches a constant value. In addition, we show that the finite-size of the system only introduces corrections of the second order or higher, which are negligible in the measured spectrum. On the other hand, the finite resolution of the measurement may replace the true Luttinger power law by trivial laws, i.e. the exponents of the momentum distribution and the density of states may approach one and zero respectively. These features should be taken into account to ensure that the measured power law is physical and not artifacts of measurement protocols.

\begin{figure*}
   	\begin{minipage}{0.01\textwidth}
   		\rotatebox{90}{\hspace{0.2in} $G (\mu S)$}
   	\end{minipage}
   	\begin{minipage}{0.92\textwidth}
   		\centering
   		\includegraphics[width=0.45\textwidth]{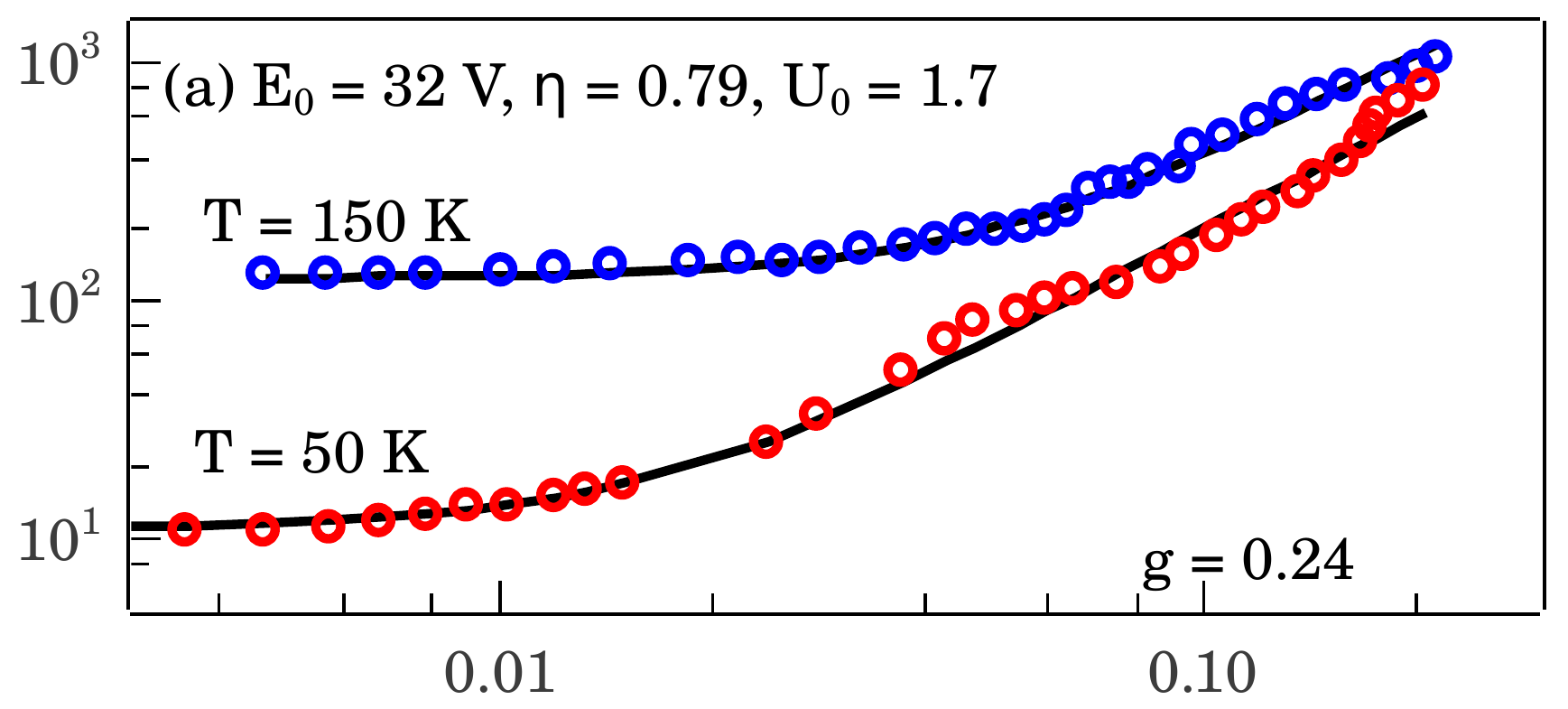}
   		\includegraphics[width=0.24\textwidth]{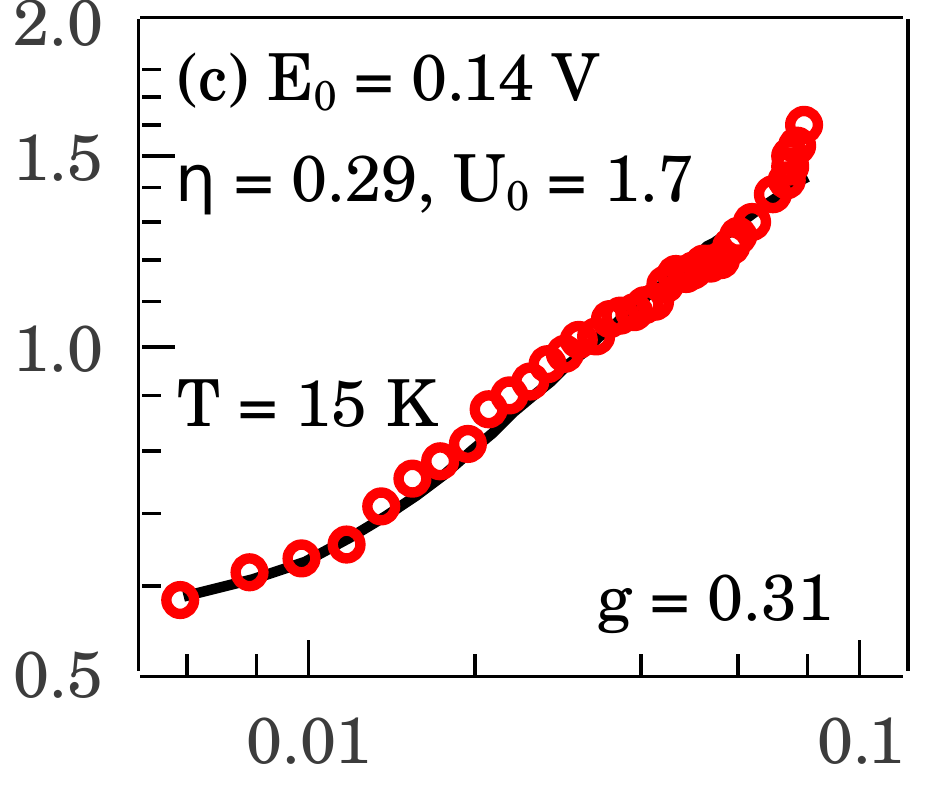}
   		\includegraphics[width=0.24\textwidth]{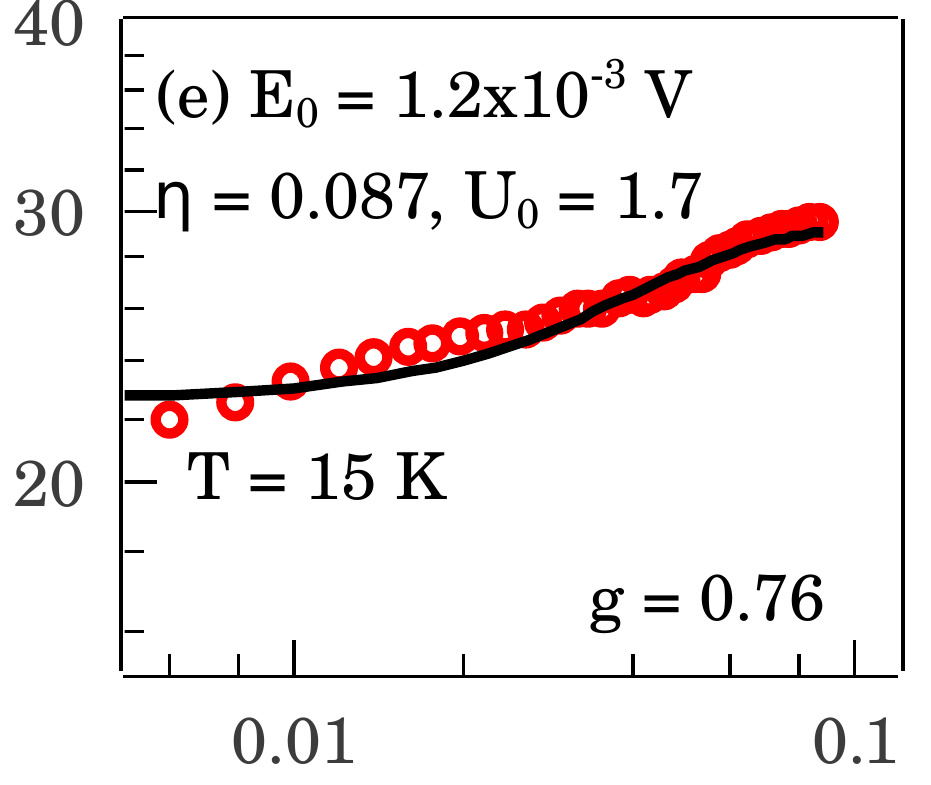}						  	 		
   		
   		\includegraphics[width=0.45\textwidth]{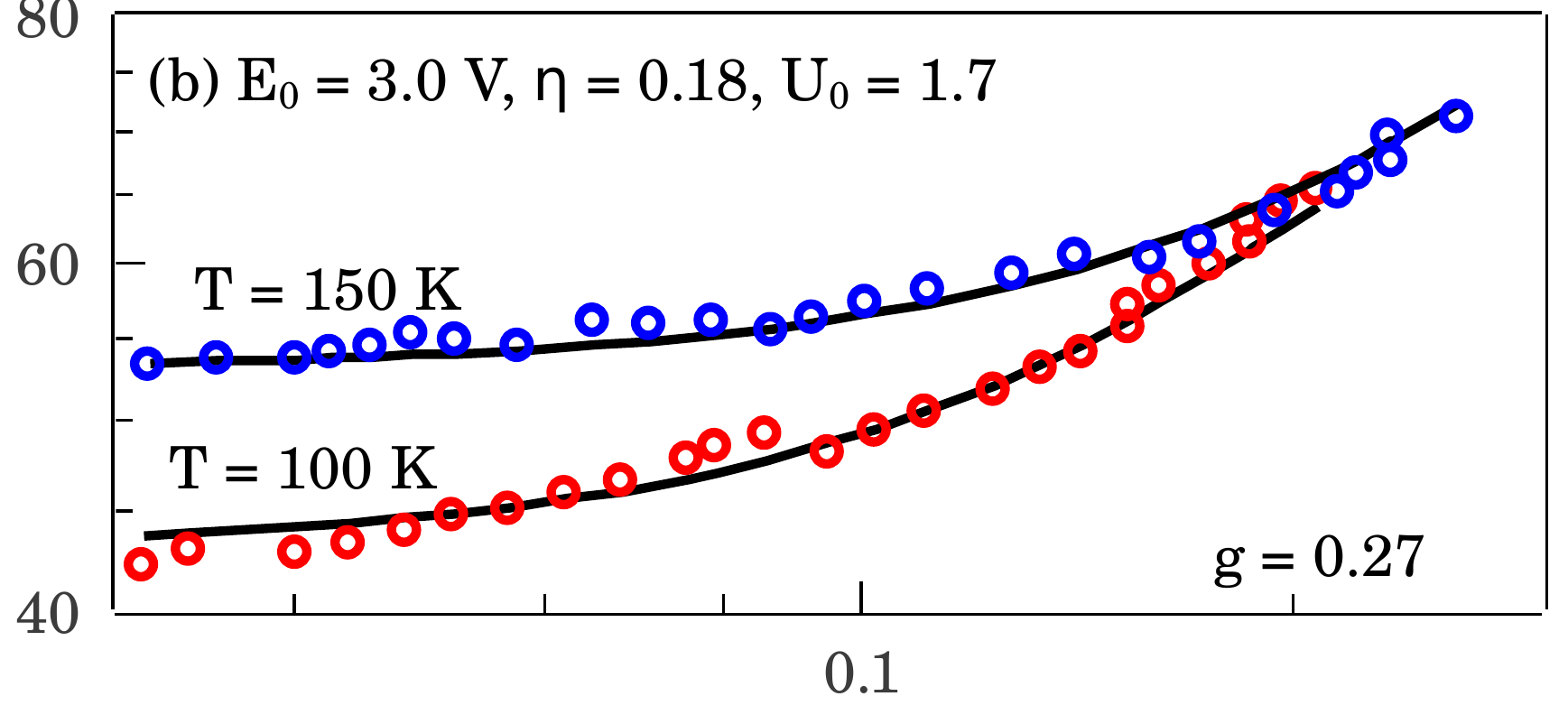}
   		\includegraphics[width=0.24\textwidth]{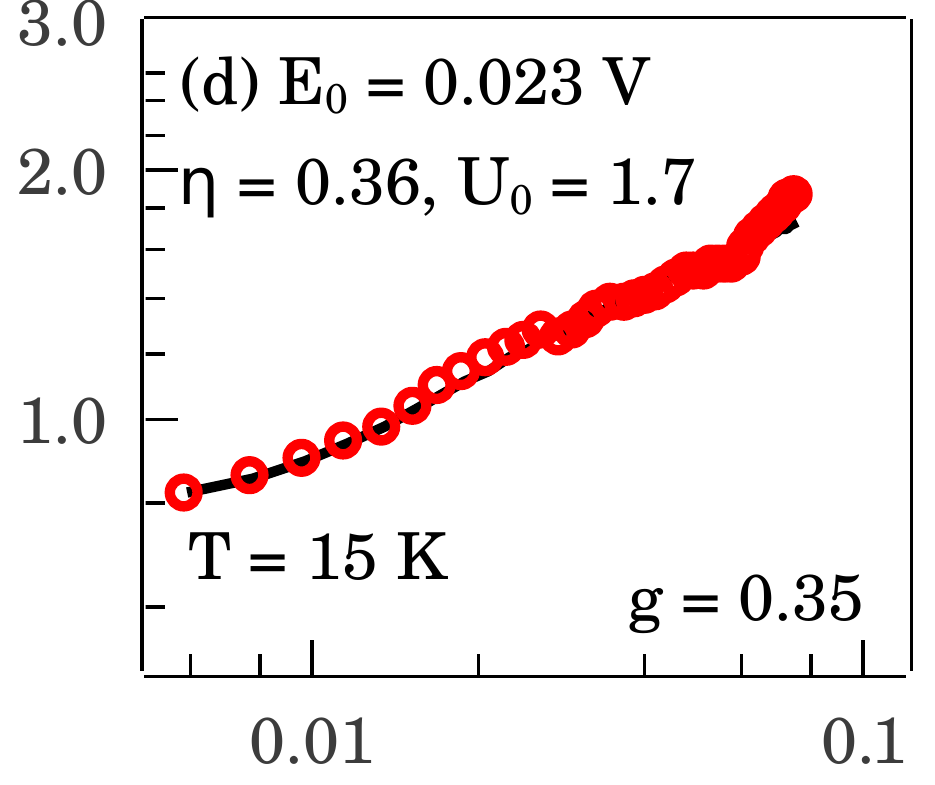}
   		\includegraphics[width=0.24\textwidth]{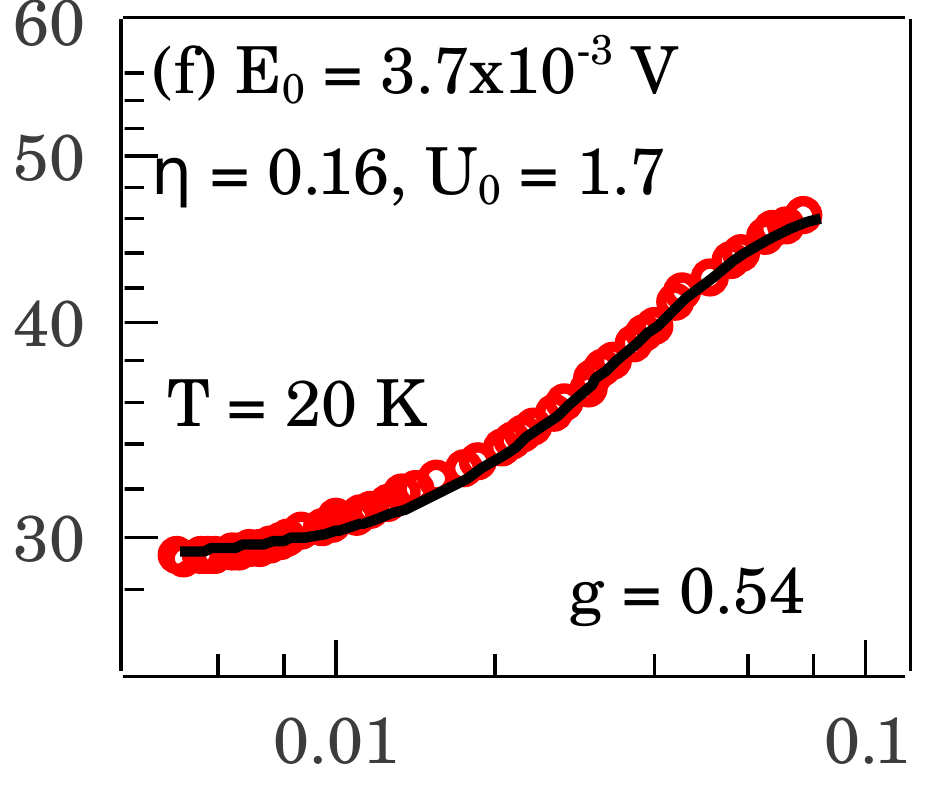}	
   		Voltage bias (V)							  	 		
   	\end{minipage}		
   	\caption{Coulomb Luttinger liquid tunneling conductance fitted to experimental data: (a) L-L boundary tunneling from Fig. 4b Ref.\cite{Yao1999}, (b) L-L bulk tunneling from Fig. 3a Ref.\cite{Postma2000}, (c) L-L bulk tunneling from Fig. 3b Ref.\cite{Zhao2018}, (d) L-L bulk tunneling from Fig. 3c Ref.\cite{Zhao2018}, (e) L-M boundary tunneling from Fig. SM4a supplement Ref.\cite{Zhao2018}, (f) L-M tunneling from Fig. SM4b supplement Ref.\cite{Zhao2018}}\label{fig6}	  		
   \end{figure*}

Having considered all factors that may affect the tunneling conductance, we finally conclude by considering (Fig.~\ref{fig6}) a comparison between theory and experiments. We focus on carbon nanotube experiments in Refs. \cite{Yao1999,Postma2000,Zhao2018}.
The Coulomb Luttinger liquid has two parameters: the interaction strength $U_0$ and the energy scale or crossover energy $E_0=v_F/d$.  Unfortunately, all the reported data have dynamic range of less than 2 orders of magnitude, hence fitting to find both $U_0$ and $E_0$ is not possible as shown earlier. (In addition, such a small dynamic range makes any distinction between long-range and short-range Luttinger liquid models essentially impossible.) However, we assume that the interaction strength is universal for all carbon nanotubes while the crossover energy scale can vary depending on the sample. Therefore, we fix the value of $U_0=1.7$ and treat $E_0$ as the fitting parameter. It is noted that the choice $U_0=1.7$ is arbitrary, as the data range is insufficient for an accurate fitting; we can also choose another $U_0$ and the values of $E_0$ will change accordingly. We also add a rescaling factor so that  $V\to \eta V$ where $\eta(<1)$ accounts for the real voltage across the tunneling contact after subtracting out the voltage drop along conductors. In Fig.~\ref{fig6}, we show the fitted parameters along with the value $g$ obtained by fitting the data to a line. This result suggests that different values of the measured interaction parameter $g$ may rise entirely from the sample-to-sample variations in the crossover scale $E_0$.  This should be taken into account in future experiments on Luttinger liquids. 

Our theoretical analysis for carbon nanotube assumes SU(2)$\times$SU(2) symmetry, which is valid under most experimental conditions. For example, in Ref.~\cite{Yao1999}, a pair of pentagon and heptagon is inserted to the hexagonal carbon lattice to create a kink that acts as a semiconductor junction between two straight nanotube segments. However, the bulk of each segment is still pristine and the valley symmetry is not broken in the bulk. The degeneracy of the system indeed simplifies the problem significantly. In principle, tunneling between multiple (more than 2) Luttinger liquids is complicated due to a large number of possible tunneling channels and possibly different behaviors in each channel \cite{Yjunction1,Yjunction2,Yjunction3,Yjunction5,Yjunction6, Yjunction7,multimode1,multimode2,multimode3,multimode4, multimode6, multimode7}. However, in a degenerate system as in our work, the symmetry of electron species enforces the same $V_0,T-$dependence on all the channels, thus effectively reducing the problem to the tunneling between only two Luttinger liquid modes. Although beyond the scope of this paper, the symmetry breaking situation (e.g. valley polarization) can be straightforwardly incorporated in the Luttinger liquid formalism. Hence, our theoretical treatment of long-range Coulomb interaction can be easily generalized to the multiple non-degenerate electron species situation. As a result, the scale dependence of the Luttinger exponent with decreasing energy, which is the signature of long-range interaction, should still manifest qualitatively although some of the quantitative details of our work will change depending on the precise details of which symmetry is broken and how it is broken.
   
  \acknowledgments{
	This work is supported by the Laboratory for Physical Sciences. A. I. acknowledges financial support from the Fulbright Foundation.}   
   
   \bibliographystyle{apsrev4-1}
   \bibliography{reference}		
\end{document}